\documentclass[10pt,twocolumn,letterpaper]{article}

\usepackage[pagenumbers]{cvpr} 

\usepackage[dvipsnames]{xcolor}

\newcommand{\latent}{\mathbf{Z}}

\makeatletter
\newcommand*{\addFileDependency}[1]{
  \typeout{(#1)}
  \@addtofilelist{#1}
  \IfFileExists{#1}{}{\typeout{No file #1.}}
}
\makeatother

\newcommand{\dlafull}{\textbf{{\color{violet}\textbf{Dream}}, {\color{teal}\textbf{Lift}}, {\color{cyan}\textbf{Animate}}\xspace{}}}
\newcommand{\dla}{\textbf{{\color{violet}\textbf{D}}{\color{teal}\textbf{L}}{\color{cyan}\textbf{A}}}}
\newcommand{\dream}{\textbf{{\color{violet}{Dream}}}}
\newcommand{\lift}{\textbf{{\color{teal}{Lift}}}}
\newcommand{\animate}{\textbf{{\color{cyan}{Animate}}}}
\definecolor{cvprblue}{rgb}{0.21,0.49,0.74}
\usepackage[pagebackref,breaklinks,colorlinks,citecolor=cvprblue]{hyperref}

\def\paperID{125} 
\def\confName{3DV\xspace}
\def\confYear{2026\xspace}

\title{Dream, Lift, Animate: \\ From Single Images to Animatable Gaussian Avatars}

\author{
  Marcel C. Buehler\\
  ETH Zurich \\
  \and
  Ye Yuan \\
  NVIDIA \\
  \and
  Xueting Li \\
  NVIDIA \\
  \and
  Yangyi Huang\\
  Chinese University of Hong Kong \\
  \and
  Koki Nagano \\
  NVIDIA \\
  \and
  Umar Iqbal \\
  NVIDIA
}

\begin{document}
\twocolumn[{%
  \maketitle
  \begin{center}
\small
\setlength{\tabcolsep}{2pt}
\newcommand{\inputheight}{3cm}
\newcommand{\imagewidth}{2cm}
\begin{tabular}{c cccccccccc}
  \includegraphics[height=\inputheight]{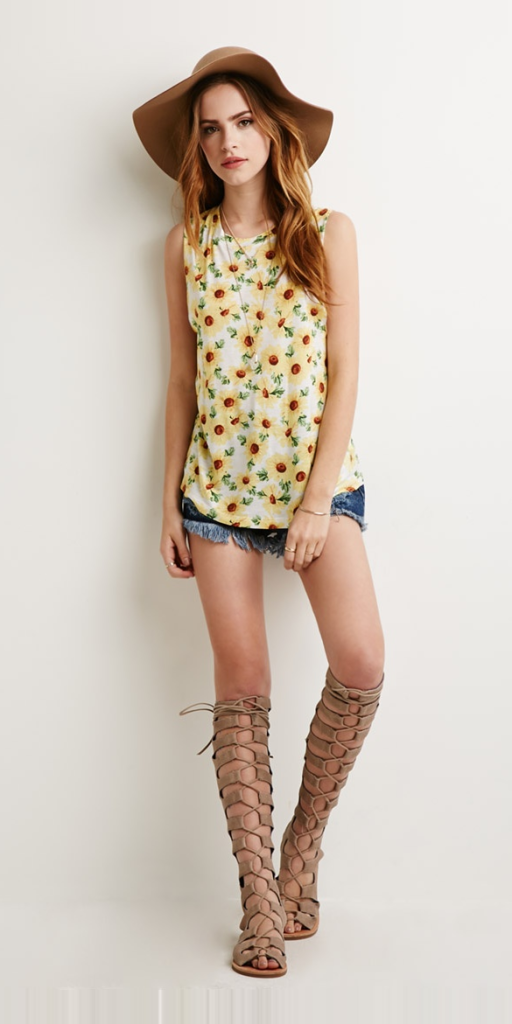}  &
  \includegraphics[height=\inputheight]{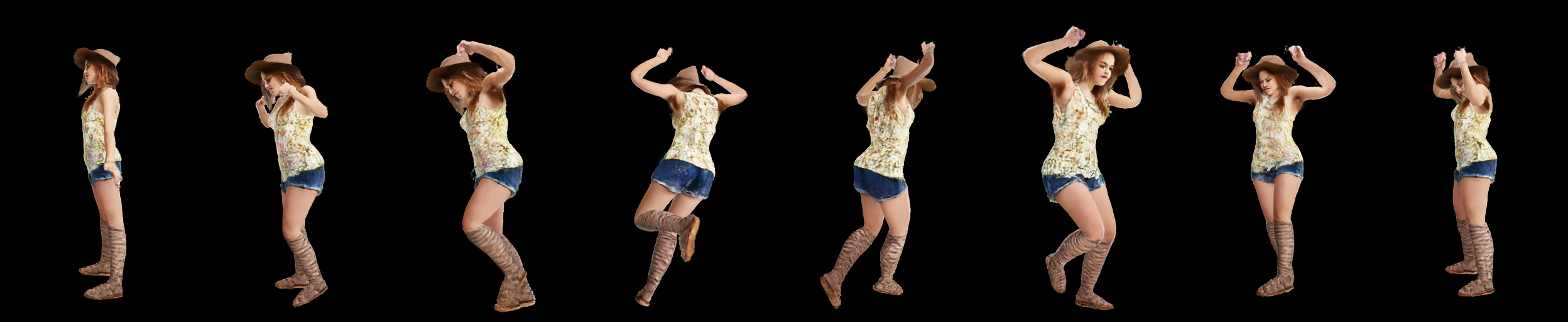} \\
  \includegraphics[height=\inputheight]{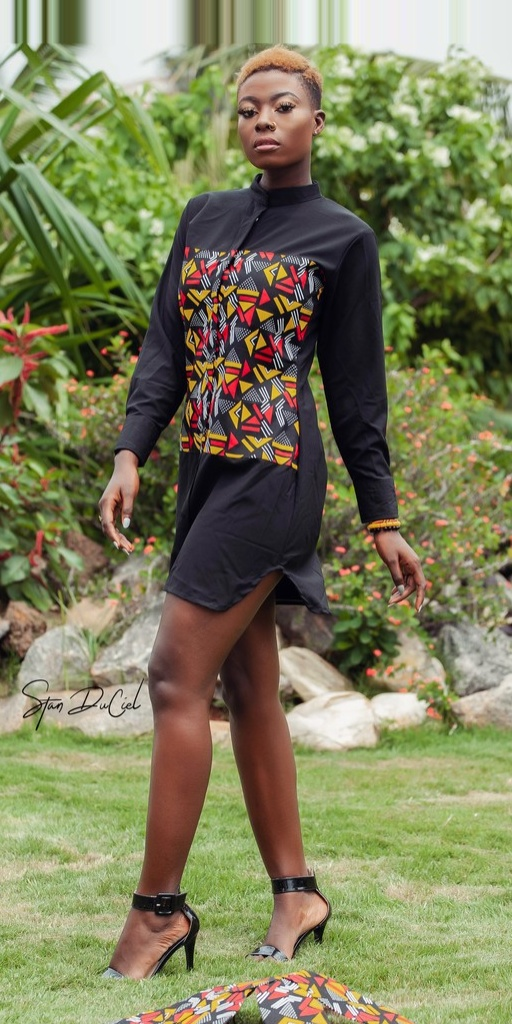}  &
  \includegraphics[height=\inputheight]{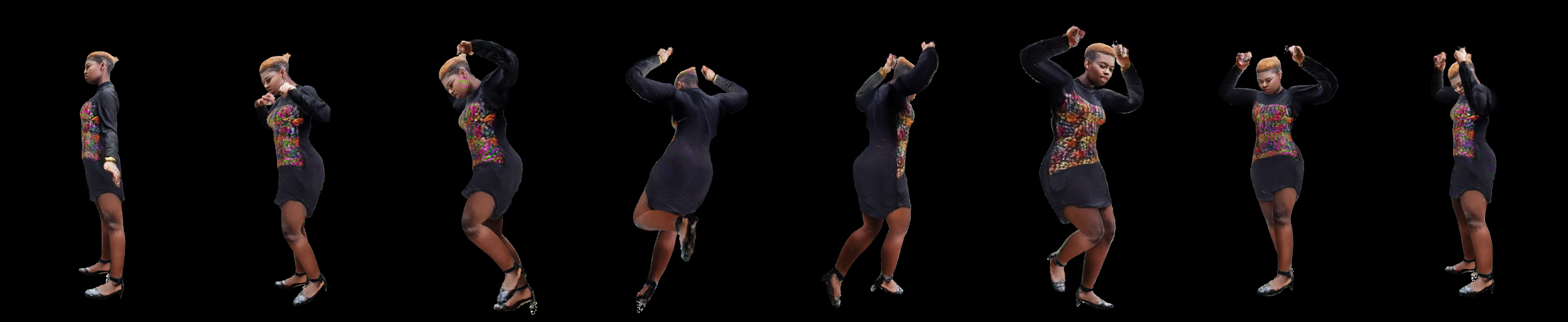} \\
  \includegraphics[height=\inputheight]{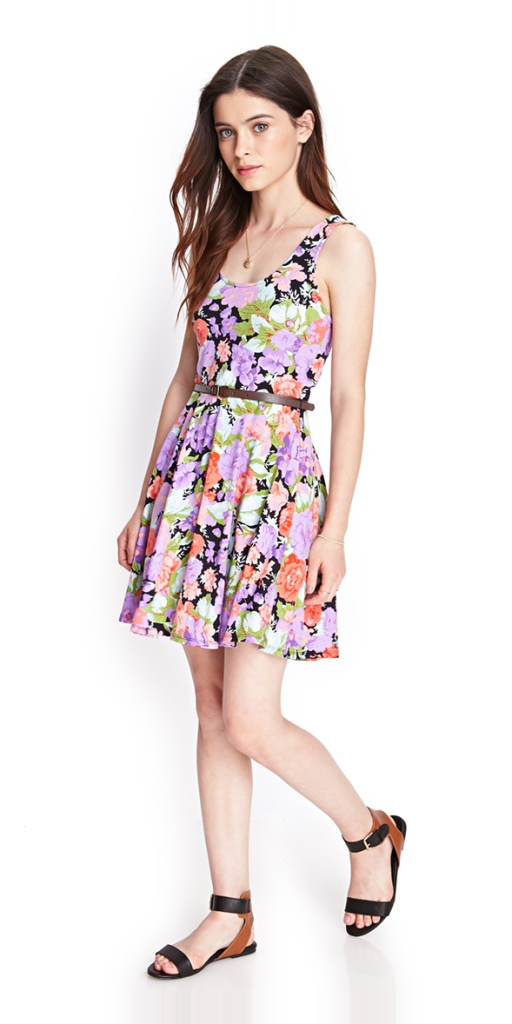}  &
  \includegraphics[height=\inputheight]{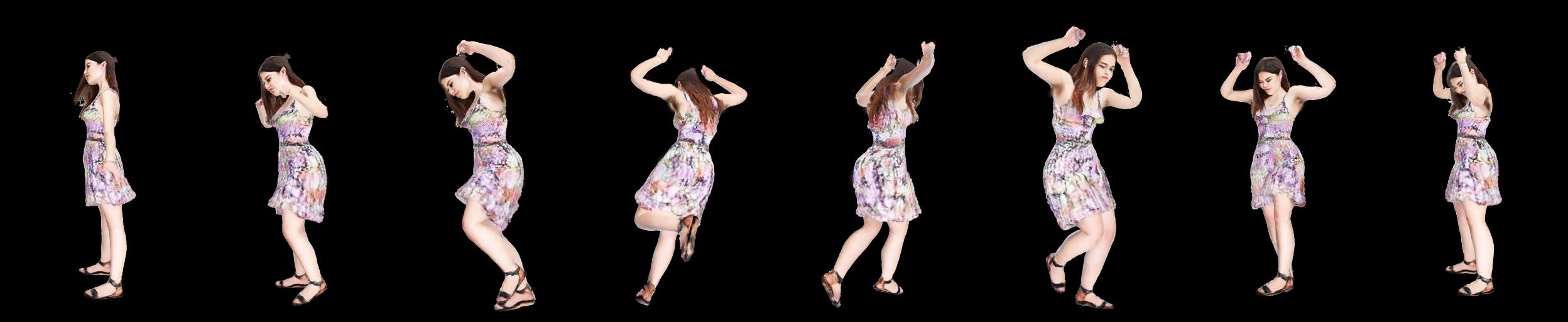} \\
  Input & Novel views and poses from a single Image \\
\end{tabular}
\captionof{figure}{\label{fig:teaser}
We propose \textbf{{\color{violet}\textbf{Dream}}, {\color{teal}\textbf{Lift}}, {\color{cyan}\textbf{Animate}}}, a novel framework to reconstruct high-fidelity, animatable 3D human avatars from a single image by generating multi-view images, lifting them to 3D Gaussians, and mapping them to a pose-aware UV space (Fig. \ref{fig:overview}). Our approach enables realistic animation and outperforms prior methods in visual quality (Fig. \ref{fig:comp_ahq_light} and Tbl. \ref{tbl:comp_ahq_light}). Watch videos and more at \small \href{https://research.nvidia.com/labs/dair/dream-lift-animate}{https://research.nvidia.com/labs/dair/dream-lift-animate}.
}
  \end{center}
}]

\maketitle

\begin{abstract}

We introduce Dream, Lift, Animate (DLA), a novel framework that reconstructs animatable 3D human avatars from a single image. This is achieved by leveraging multi-view generation, 3D Gaussian lifting, and pose-aware UV-space mapping of 3D Gaussians. Given an image, we first dream plausible multi-views using a video diffusion model, capturing rich geometric and appearance details. These views are then lifted into unstructured 3D Gaussians. To enable animation, we propose a transformer-based encoder that models global spatial relationships and projects these Gaussians into a structured latent representation aligned with the UV space of a parametric body model. This latent code is decoded into UV-space Gaussians that can be animated via body-driven deformation and rendered conditioned on pose and viewpoint. By anchoring Gaussians to the UV manifold, our method ensures consistency during animation while preserving fine visual details. DLA enables real-time rendering and intuitive editing without requiring post-processing. Our method outperforms state-of-the-art approaches on the ActorsHQ and 4D-Dress datasets in both perceptual quality and photometric accuracy. By combining the generative strengths of video diffusion models with a pose-aware UV-space Gaussian mapping, DLA bridges the gap between unstructured 3D representations and high-fidelity, animation-ready avatars.

\end{abstract}
\section{Introduction}
\begin{figure*}[t]
    \centering
  \includegraphics[width=\textwidth]{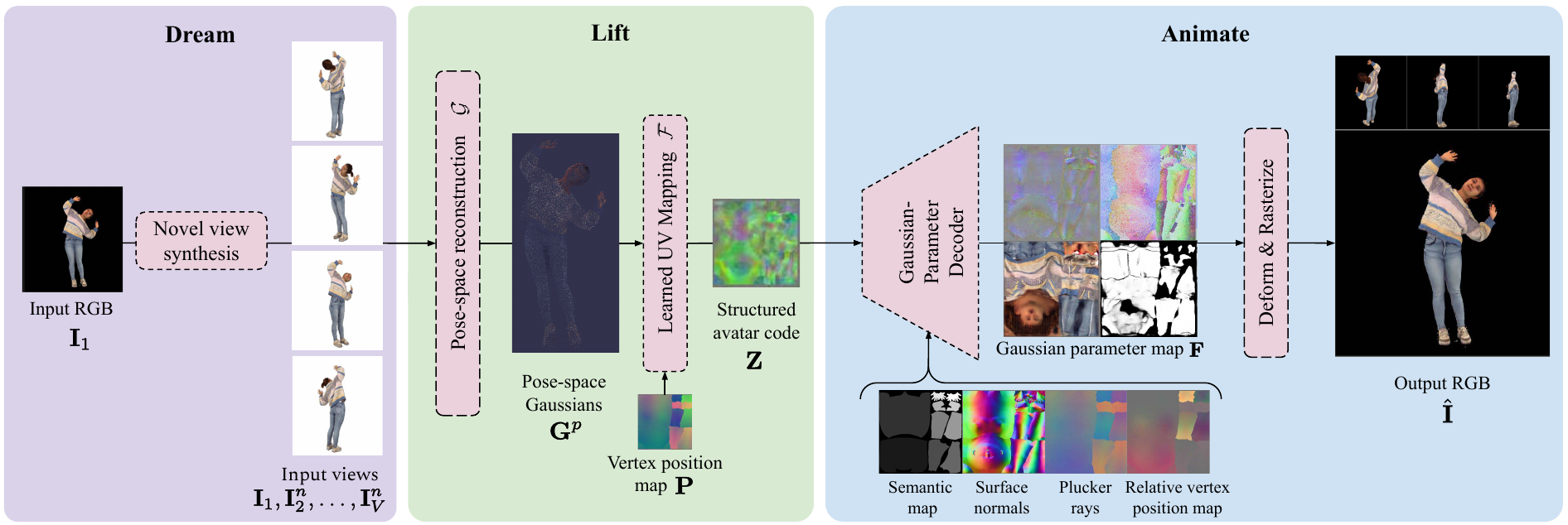}
    \caption{\label{fig:overview}Overview of the proposed \dlafull ~(\dla) framework for reconstructing animatable 3D human avatars from a single image. In the {\color{violet}\textbf{Dream}} stage (Sec. \ref{sec:dream}), we synthesize novel views from the input using a diffusion-based generator. In the {\color{teal}\textbf{Lift}} stage (Sec. \ref{sec:lift} and Fig. \ref{fig:encoder}), we project the multi-view images into a set of unstructured 3D Gaussians in the \emph{pose space}  using a learned Gaussian reconstruction model $\mathcal{G}$. Subsequently, we learn a transformer encoder $\mathcal{F}$ to map 3D Gaussians to a structured latent code $\mathbf{Z}$ in the UV space of a parametric body model. In the {\color{cyan}\textbf{Animate}} stage (Sec. \ref{sec:animate} and Fig. \ref{fig:gpd}), we decode the avatar code into a pose- and view-aware Gaussian parameter map $\mathbf{F}$. This structured representation enables realistic animation and rendering via deformation with a body model.\vspace{-1em}}
   
\end{figure*}

Creating photorealistic 3D human avatars from monocular RGB images remains a fundamental and challenging problem in computer vision and graphics, with wide-ranging applications in gaming, telepresence, virtual try-on, and digital content creation. Achieving high-quality avatar reconstruction from a single image requires addressing three interrelated challenges. First, missing appearance details arising from self-occlusions or limited camera viewpoints must be plausibly and photorealistically hallucinated. Second, sparse 2D information from the input image must be reliably lifted into a geometrically coherent and accurate 3D representation. Third, and most critically, the resulting avatars must be readily animatable, enabling realistic and artifact-free motion synthesis under novel poses and viewpoints while maintaining consistent geometry, texture, and view-dependent effects.

Existing methods typically fall short in simultaneously addressing these challenges. Recent approaches~\cite{ho2024sith, xue2024human3diffusion,he2024magicman} leverage the powerful generative capabilities of multi-view diffusion models~\cite{Rombach2022StableDiffusion} to hallucinate missing appearance information; however, these models inherently produce inconsistencies across generated views, leading to avatars that appear overly smoothed, lack critical details, or introduce visually jarring artifacts when animated. Conversely, methods relying on template-based rigging approaches, such as automatically transferring skinning weights from canonical human body models (e.g., SMPL~\cite{loper2015smpl}), typically require careful fitting procedures \cite{zhang2023sifu,huang2022elicit,huang2024tech,svitov2023dinar}. While powerful, these methods can still encounter challenges when handling non-standard poses, complex clothing, and significant self-occlusions frequently encountered in unconstrained, real-world scenarios, thus affecting their robustness and broader applicability.

To overcome these limitations, we propose \textbf{\textbf{{\color{violet}\textbf{Dream}}, {\color{teal}\textbf{Lift}}, {\color{cyan}\textbf{Animate}}} (\textbf{{\color{violet}\textbf{D}}{\color{teal}\textbf{L}}{\color{cyan}\textbf{A}}})}, a novel framework designed specifically for reconstructing high-quality, animatable 3D human avatars from a single image. Our method addresses the core reconstruction problem by decomposing it into three complementary stages, each carefully designed to handle limitations inherent in the preceding steps. Specifically, we first leverage a pretrained video diffusion model \cite{wang2024unianimate} to \dream{} plausible multi-view observations of the subject, effectively hallucinating realistic geometric and appearance details even in regions unseen from the input viewpoint. Despite the significant progress made by recent diffusion-based approaches, their inherent view-to-view inconsistencies prevent direct avatar reconstruction~\cite{xue2024human3diffusion}. To resolve this, we next \lift{} these generated views into an intermediate, unstructured 3D representation based on 3D Gaussian primitives~\cite{kerbl20233d} in the input pose space, effectively aggregating appearance and geometry cues from the multi-view observations. Subsequently, to enable animation, we map these unstructured Gaussians into a structured UV-space representation. For this, we use a transformer-based encoder~\cite{zhang3dshape2vecset} that models global spatial relationships among the unstructured 3D Gaussians and projects them into a structured latent avatar representation, which aligns with the UV space of the SMPL-X body model~\cite{pavlakos2019smplx}. This learned mapping also allows our model to reconcile inconsistencies introduced during the initial multi-view generation seamlessly. 
Finally, in the \animate{} stage, we present a pose- and view-aware Gaussian parameter decoder that converts this latent avatar representation into a UV-space map of Gaussian parameters, enabling spatially coherent and high-fidelity animation through linear blend skinning-based Gaussian deformation.

We demonstrate the advantages of our proposed approach through extensive experiments on the challenging ActorsHQ~\cite{isik2023humanrf}, 4D-Dress ~\cite{wang20244ddress}, and SHHQ \cite{fu2022styleganhuman} datasets. 
Our method achieves state-of-the-art performance in terms of photometric accuracy, perceptual quality, and articulation realism.

In summary, the core contributions of our work are:

\begin{itemize}
\item A novel framework (\textbf{DLA}) for reconstructing animatable 3D human avatars from a single monocular RGB image, overcoming the inherent limitations of diffusion-based multi-view generation and template-based rigging approaches.
\item A learned transformer-based encoder coupled with a pose- and view-conditioned Gaussian Parameter Decoder that maps unstructured 3D Gaussians into a structured UV-space representation, enabling high-fidelity animation.
\item Extensive experiments and ablations demonstrating the effectiveness and versatility of our method, including one-shot reconstruction, realistic animation, and editing.
\end{itemize}
\section{Related Work}

\noindent\textbf{One-Shot Human Reconstruction.}
Reconstructing 3D human avatars from a single image has been explored using both mesh-based and volumetric representations. Mesh-based approaches, including PiFU(-HD) \cite{saito2019pifu, saito2020pifuhd} and its extensions \cite{ho2024sith, zhang2023sifu, alldieck2022phorhum, sengupta2024diffhuman, zheng2020pamir, xiu2023econ, xiu2023icon, zhang2023globalcorrelated}, typically rely on implicit surfaces or hybrid models~\cite{shen2021dmtet, huang2024tech, yang2024hilo} to reconstruct geometry and texture. While some decouple albedo and shading~\cite{sengupta2024diffhuman, alldieck2022phorhum, corona2023s3f}, mesh-based methods often struggle to model realistic view-dependent effects.
Volumetric representations address these limitations. Works based on NeRFs~\cite{hu2023sherf, huang2022elicit, weng_humannerf_2022_cvpr, gao2024contex, buhler2023preface, buehler2024cafca} achieve high fidelity but remain slow to render. Recent methods adopt 3D Gaussian Splatting~(3DGS) \cite{kerbl20233d} for efficient, real-time rendering. Human-focused variants \cite{xue2024human3diffusion, tang2025lgm, pan2024humansplat, liu2024humanvdm,chen2024generalizable} use diffusion-generated multiviews~\cite{voleti2024sv3d} to supervise Gaussian reconstruction. However, these typically model static pose-space geometry and lack support for animation under novel poses.

\noindent\textbf{Animatable Avatars.}
To enable animation, early works like ARCH~\cite{huang2020arch, he2021arch++} use canonicalization with LBS deformation, extended by feedforward models~\cite{hu2023sherf, peng2024charactergen}. These methods, however, depend on accurate body registration, which is often challenging for complex poses and clothing. IDOL~\cite{zhuang2024idolinstantphotorealistic3d} and LHM~\cite{qiu2025LHM} (concurrent work) directly predict Gaussian splats on the SMPL-X template to support animation. However, their deformation module does not consider pose-dependent effects. In contrast, our approach conditions the Gaussian parameters on the target pose, enabling pose-dependent effects. In addition, our approach is substantially more lightweight because it decomposes the task into more tractable subproblems of multiview hallucination, 3D lifting, and UV-space mapping. While IDOL and LHM train on multiple nodes (32 NVIDIA A100/H100 GPUs), our model only requires a single node (8 NVIDIA A100 GPUs). Our lightweight design enables modularity and better handling of pose- and view-dependent appearance effects essential for animation.

\noindent\textbf{Generative Priors.}
Generative models such as GANs~\cite{gan} and diffusion models~\cite{Rombach2022StableDiffusion, lipmanflow} have also been used as strong priors for ill-posed problems such as single-image human reconstruction~\cite{ho2024sith, zhang2023sifu, albahar2023humansgd, pan2024humansplat, gao2024contex, zhang2024humanref, li2024pshuman, lu2025gas, huang2025adahuman}.
Recent methods apply diffusion models for view synthesis and geometry prediction.
Human3Diffusion~\cite{xue2024human3diffusion} jointly trains a 3DGS generator with a multiview diffusion model, but remains limited by low resolution ($256 \times 256$) and a lack of animation support. SiTH~\cite{ho2024sith} and HumanSplat~\cite{pan2024humansplat} synthesize unseen views using single- or multi-view diffusion, while SIFU~\cite{zhang2023sifu} uses a transformer to predict geometry and refines textures with diffusion. Our method builds on both paradigms, i.e., we leverage a pretrained video diffusion model \cite{wang2024unianimate} to synthesize unseen views of the person, and we use a PatchGAN \cite{isola2017image} to achieve higher realism and high-frequency details.

\noindent\textbf{Generating 3D Avatars as Latents.}
Learning generative models of humans in a structured latent space enables controllable synthesis and supports downstream tasks such as few-shot reconstruction, inpainting, and editing. 
Early 3D methods map latent codes to implicit fields~\cite{dong2023ag3d, bergman2022gnarf, hong2023evad} or SDFs~\cite{noguchi2022unsupervised, zhang2023getavatar, deepsdf}, often using tri-plane features and adversarial supervision. To improve pose control and spatial detail, recent works incorporate body  priors~\cite{loper2015smpl, pavlakos2019smplx} with structured latents~\cite{hu2024structldm, chen2023primdiffusion, abdal2023gaussian, kolotouros2024avatarpopup, hong2023evad, men2024en3d}. 
However, these methods still fall short in texture realism and often rely on optimization-based inversion~\cite{10.1145/3658162, roich2021pivotal} for image/text conditioning.
In contrast, our method directly maps an input image to a compact, UV-aligned avatar latent that naturally supports animation, editing, and even interpolation.

\section{Method}
\label{sec:method}

Fig.~\ref{fig:overview} provides a high-level overview of our proposed framework for reconstructing animatable 3D human avatars from a single image. Given the input image $\mathbf{I}_1 \in \mathbb{R}^{H_I \times W_I \times C_I}$, our method proceeds in three stages: \dlafull. In the first stage, \dream, we employ a video diffusion model~\cite{wang2024unianimate} to generate plausible multi-view images ${\mathbf{I}_2^n, \dots, \mathbf{I}_V^n}$ from the input, addressing self-occlusions and incomplete viewpoints. Although visually compelling, these generated views often exhibit inconsistencies. Thus, in the second stage, \lift, we lift these multi-view images into a coherent set of unstructured pose-space 3D Gaussians ${\mathbf{G}_1^p, \dots, \mathbf{G}_K^p}$ and use a novel transformer-based encoder to transform them into an animation-friendly structured latent representation $\mathbf{Z}$, which is aligned with the UV space of the SMPL-X body model~\cite{pavlakos2019smplx}. 
Finally, in the third stage, \animate, a Gaussian Parameter Decoder predicts a Gaussian parameter map $\mathbf{F}$ in the UV space from the latent code $\mathbf{Z}$, target pose, and viewpoint conditions. We can then sample structured Gaussians $\mathbf{G}_1^s, \ldots, \mathbf{G}_N^s$ from the parameter map. These Gaussians can be readily animated via linear blend skinning (LBS) and rendered in real-time:
$\mathbf{\hat I} = \mathcal{R}(\{\mathbf{G}_1^s, \ldots, \mathbf{G}_N^s\}, \Theta, \mathbf{\pi}),$
where $\mathcal{R}$ is a rendering function that deforms the structured Gaussians $\mathbf{G}_1^s, \ldots, \mathbf{G}_N^s$ according to target pose $\Theta$ using SMPL-X linear blend skinning, and rasterizes them with camera parameters $\mathbf{\pi}$.

\begin{figure*}[t]
    \centering
  \includegraphics[width=\textwidth]{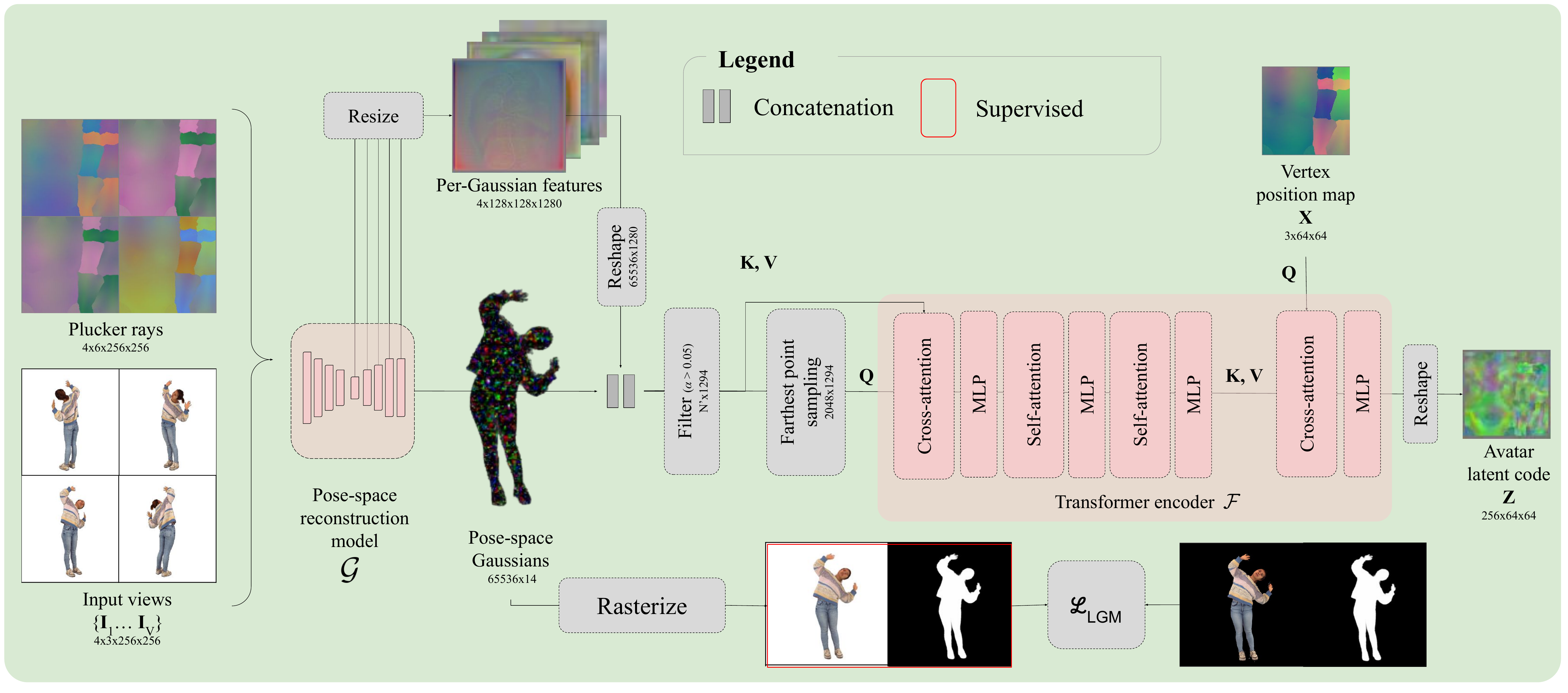}
    \caption{\label{fig:encoder}\emph{\lift ~from multiview images to an avatar latent code}. The pose-space reconstruction model produces pixel-aligned Gaussian parameters with corresponding feature maps, denoted \emph{pose-space Gaussians} and \emph{per-Gaussian features}. The Gaussians are filtered and subsampled to construct a compact Gaussian feature $\mathbf{X}$ with 2048 Gaussians. These inputs are further processed by a transformer. Specifically, the compact Gaussian feature serves as context (key $\mathbf{K}$ and value $\mathbf{V}$) to a cross-attention layer with queries $\mathbf{Q}$ being the positionally encoded vertex position map $\mathbf{P}$. Finally, the output is reshaped and yields the avatar latent code $\mathbf{Z}$. This figure omits linear projections, skip-connections, and positional encoding for improved readability.
    }
\end{figure*}

\subsection{Dream: Multi-view Generation} 
\label{sec:dream}
As illustrated in Fig.~\ref{fig:overview}, our method begins by generating a set of multi-view images consisting of the original input view $\mathbf{I}_1$ and a set of novel views $\mathbf{I}_2^n, \ldots, \mathbf{I}_V^n$. During training, these views are sourced directly from multi-view datasets with known camera calibrations. At inference time, however, we synthesize novel views using a ControlNet-guided variant~\cite{wang2024unianimate} of a video diffusion model~\cite{wan2025}. Specifically, we first estimate the SMPL-X parameters from the input image \cite{he2024magicman} and render 2D skeletal poses of the predicted mesh from virtual cameras placed around a 360-degree azimuth. These projected poses serve as control signals to guide the diffusion model in generating photorealistic images from novel viewpoints. While this approach enables hallucination of previously unseen regions and recovers occluded appearance details, the resulting views may still exhibit 3D inconsistencies due to artifacts inherent to the generative diffusion process. Please see the supplementary material for more details.

\subsection{Lift: Unstructured Gaussian and Latent Avatar Code Generation}
\label{sec:lift}

Fig. \ref{fig:encoder} illustrates how we lift the multiview observations into a unified 3D avatar representation in two sequential sub-stages. First, we reconstruct per-view 3D Gaussians in the input pose space. Second, we map these unstructured 3D Gaussians into a structured latent avatar code in the UV space of the SMPL-X body model, which is more amenable to animation.

\noindent\textbf{Unstructured Gaussians Reconstruction.}
We employ a pose-space reconstruction model $\mathcal{G}$ to map the multiview images into a set of pixel-aligned 3D Gaussians. The model uses a U-Net-based architecture augmented with cross-view self-attention, following the design of the Large Gaussian Model (LGM)~\cite{tang2025lgm}. The resulting Gaussians from all views are then fused into a single set of unstructured 3D Gaussians. Reconstructing Gaussians directly in the input pose space is a tractable and effective strategy, as it avoids the highly non-linear mappings required to directly predict avatar geometry in canonical or UV coordinates. This design choice allows our method to faithfully recover rich appearance and geometric details from the input views while remaining robust to inconsistencies.

\noindent\textbf{Latent Avatar Code Generation.}
While informative, these unstructured Gaussians are not directly amenable to animation because they lack a consistent topology and structure. To address this, we introduce a transformer-based encoder $\mathcal{F}$ that converts the unstructured Gaussians into a structured latent avatar code $\latent$. The latent code is aligned with the UV space of the SMPL-X model~\cite{pavlakos2019smplx}, which supports expressive deformation and animation via linear blend skinning (LBS). As illustrated in Fig.~\ref{fig:encoder}, we first extract per-Gaussian features by concatenating each Gaussian’s raw parameters (e.g., color, position) with intermediate U-Net activations and linearly project them to an embedding space. These features are pixel-aligned by construction, owing to the design of the U-Net. The per-Gaussian features are then filtered and downsampled with farthest-point sampling \cite{eldar1997farthest} into a more compact Gaussian feature $\mathbf{X} \in \mathbb{R}^{P \times C_p}$ where $P$ are the number of filtered Gaussians and $C_p$ is the combined dimensionality of the Gaussian parameters and U-Net features.

To associate this representation with the SMPL-X UV space, we use a UV-space vertex position map $\mathbf{P} \in \mathbb{R}^{3 \times H_p \times W_p}$ generated by deforming the SMPL-X mesh with the input pose $\Theta_I$ and rasterizing the mesh vertex coordinates into UV space. This position map is positionally encoded and used as queries in a cross-attention layer, where the Gaussian features $\mathbf{X}$ serve as keys and values. The result is a compact, spatially-structured latent code $\latent$ aligned with the UV manifold of the SMPL-X body.

This design provides several critical advantages. First, reconstructing Gaussians in the input pose space and lifting them into UV space decouples local appearance estimation from structural reasoning, simplifying both tasks. Second, the feed-forward nature of our lifting pipeline supports fast inference and enables end-to-end differentiable training. Finally, by leveraging a learned reconstruction model and a structured UV mapping, our method effectively absorbs view inconsistencies and artifacts in the generated multiview inputs, producing a coherent and animatable intermediate representation.

\subsection{Animate: Structured Gaussian Generation and Deformation}

\begin{figure*}[t]
    \centering
  \includegraphics[width=\textwidth, trim={0 0 0 0mm},clip]{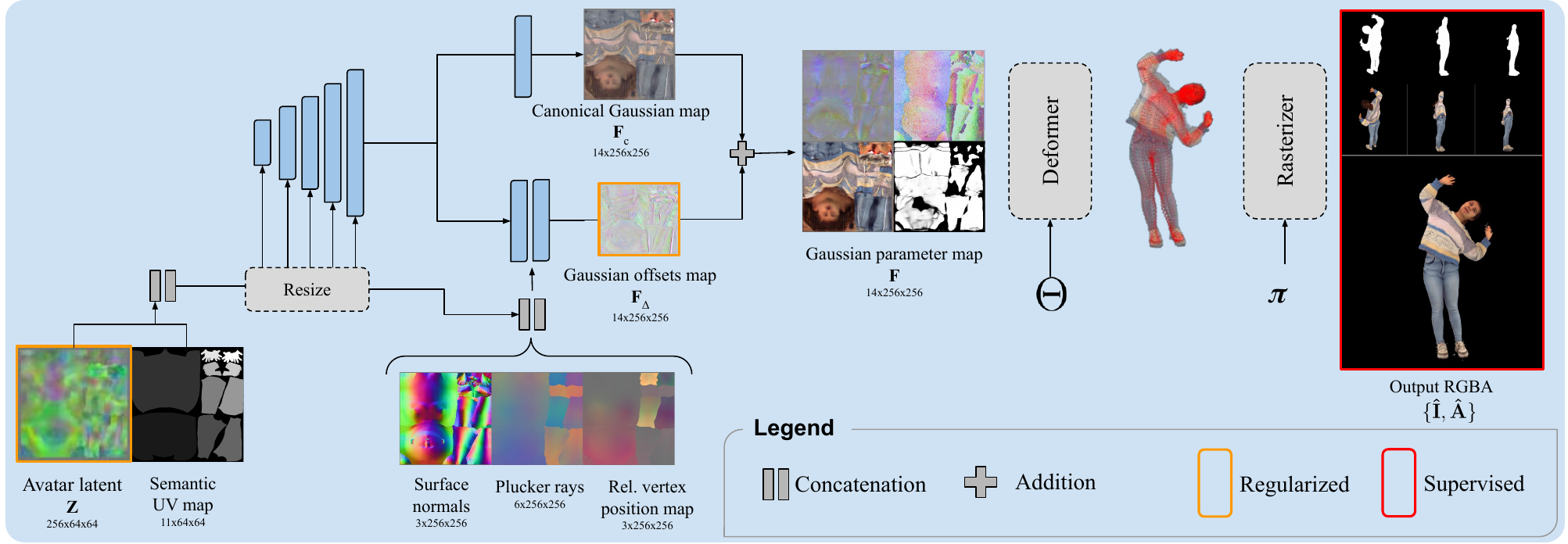}
    \caption{\label{fig:gpd}\animate. ~The Gaussian Parameter Decoder (GPD, Sec. \ref{sec:gpd}) maps a UV-aligned latent $\mathbf{Z}$ to an animatable 3D Gaussian representation. The GPD upsamples the avatar latent code $\mathbf{Z}$ and produces two output maps: a \emph{canonical} Gaussian map $\mathbf{F}_c$ and an \emph{offset} map $\mathbf{F}_\Delta$. The offset map $\mathbf{F}_\Delta$ adds pose- and view-dependent offsets to the canonical Gaussian $\mathbf{F}_c$, enabling pose- and view-dependent effects. Given a pose $\Theta$ and camera $\pi$, the Gaussians are deformed with linear blend skinning \cite{pavlakos2019smplx} and rasterized to an RGBA image \cite{kerbl20233d}.\vspace{-1em}}
\end{figure*}
\label{sec:animate}
\noindent\textbf{Gaussian Parameter Decoder.}
\label{sec:gpg}
Building upon the UV-aligned latent avatar code $\latent$ produced in the previous stage, we now decode this representation into a fully animatable Gaussian-based avatar.  As shown in Fig.~\ref{fig:gpd}, the Gaussian Parameter Decoder (GPD) maps the avatar latent code $\latent$ to a UV space Gaussian parameter map, from which structured 3D Gaussians can be sampled. Since convolutional networks greatly benefit from pixel-aligned input features, we design the GPD as a spatially-adaptive ConvNet \cite{park2019SPADE,Zhu_2020_CVPR}. The GPD is conditioned on the UV-aligned latent $\latent$ and a one-hot encoded UV segmentation map \cite{pavlakos2019smplx}. We condition with spatially-adaptive normalization~\cite{park2019SPADE} because pixelwise normalization has been shown to provide great results for generating images that are aligned to semantic maps \cite{park2019SPADE,Zhu_2020_CVPR}. 
At the highest resolution, the GPD branches out into a \emph{canonical} branch and a \emph{pose- and view-dependent} branch. The pose- and view-dependent branch receives additional inputs with information about the target geometry (surface normals), viewpoint (Plucker rays \cite{jia2020plucker}), and pose (relative vertex position map w.r.t. to the neutral pose). These inputs are created by instantiating the SMPL-X body model with the given target pose and rasterizing its relative vertex location, normals, and plucker rays to the UV space. The inputs are again encoded via spatially adaptive normalization and condition the pose- and view-dependent branch. The two branches are merged with an addition. The output is a UV space Gaussian parameter map $\mathbf{F} \in \mathbb{R}^{H_G \times W_G \times 14}$, which contains the 3D Gaussian parameters, including alphas, colors, as well as positions, rotations and scales defined in the UV tangent space. The two-branch design enables fast inference by caching the canonical branch. Please refer to Sec. \ref{sec:applications} for more information.

\noindent\textbf{Structured Gaussian Generation and Rendering.}
\label{sec:gpd}
We generate the structured 3D Gaussians $\{\mathbf{G}_1^s, \ldots, \mathbf{G}_N^s\}$ by sampling the UV space Gaussian parameter map $\mathbf{F}$ at uniform locations in the UV space and on the mesh surface. Each Gaussian $\mathbf{G}_i^s$ is defined in the UV tangent coordinates and contains the position $\Delta_i^\text{xyz}$, rotation $\Delta_i^\text{R}$, scale $\Delta_i^\text{s}$, alpha $\alpha_i$, and color $\mathbf{c}_i$. To obtain the Gaussians in the world space, we deform the SMPL-X body mesh using the target pose $\Theta$ via linear blend skinning, which yields the UV tangent space coordinates $(T_i, R_i, S_i)$, which consist of position $T_i$, rotation $R_i$, and scale $S_i$. We obtain the final position, rotation, and scale of each Gaussian by 
\begin{align}
    \mathbf{G}_i^\text{xyz} = T_i + R_i \Delta_i^\text{xyz}, \quad \mathbf{G}_i^\text{R} = R_i \Delta_i^\text{R}, \quad \mathbf{G}_i^\text{s} = S_i \Delta_i^\text{s}
\end{align}
and rasterize the avatar using Gaussian Splatting~\cite{kerbl20233d}.

\subsection{Training}
Training presents significant challenges as the model must learn a complex mapping from 2D images through an intermediate UV space to a 3D Gaussian parameter representation. We provide the model with ground truth inputs with 90-degree yaw angle differences and task the model with reconstructing novel viewpoints. Given the Gaussian avatar produced by GPD (Sec.~\ref{sec:gpd}), we render it using the ground-truth SMPL-X parameters and compute the following losses:
\begin{align}
\centering
        \mathcal{L}_\text{GPD} &=   \lambda_\text{L1} \mathcal{L}_\text{L1} 
                 + \lambda_\text{VGG} \mathcal{L}_\text{VGG} \\
                & + \lambda_\text{Mask} \mathcal{L}_\text{Mask} 
                 + \lambda_\text{GAN} \mathcal{L}_\text{GAN} \\
                 & + \lambda_\text{KL} \mathcal{L}_\text{KL} 
                 + \lambda_\text{C} \mathcal{L}_\text{C}, 
\end{align} 
where the L1 reconstruction loss $\mathcal{L}_\text{L1} = \lVert \mathbf{I}^* - \mathbf{\hat I} \rVert_1$ measures the absolute difference between the rendered output and ground truth image. For mask supervision, we compute $\mathcal{L}_\text{Mask} = \lVert \mathbf{M}^* - \mathbf{\hat A} \rVert_1$, which enforces consistency between the alpha maps from Gaussian splatting $\mathbf{\hat A}$ and the ground truth masks $\mathbf{M}^*$. To enhance perceptual quality, we incorporate a GAN loss $\mathcal{L}_\text{GAN}$ using a Patch Discriminator \cite{isola2017image} with least squares optimization. The perceptual VGG loss $\mathcal{L}_\text{VGG}$ leverages an AlexNet \cite{krizhevsky2017imagenet} backbone with features masked by the ground truth mask $\mathbf{M}^*$. For regularization, we apply a Kullback-Leibler Divergence term $\mathcal{L}_\text{KL}$ to constrain the avatar latent maps, and $\mathcal{L}_\text{C} = \lVert \mathbf{F}_\Delta \rVert_2$ encourages minimal offsets in the Gaussian offsets map.

We also provide intermediate supervision to the pose-space reconstruction model $\mathcal{G}$ (Sec.~\ref{sec:lift}). Since the unstructured reconstructed Gaussians are already in pose-space, we splat them and apply the following losses: 
\begin{equation}
\centering
        \mathcal{L}_\text{UG} = \lambda_\text{VGG}^\text{UG} \mathcal{L}_\text{VGG}^\text{UG} + \lambda_\text{Mask}^\text{UG} \mathcal{L}_\text{Mask}^\text{UG},
\end{equation}
where $\mathcal{L}_\text{VGG}^\text{UG}$ and $\mathcal{L}_\text{Mask}^\text{UG}$ are the VGG loss and mask loss computed on the reconstructed images, respectively. The total loss then becomes: $\mathcal{L} = \mathcal{L}_\text{GPD} + \mathcal{L}_\text{UG}$. 
Fig.~\ref{fig:encoder} and Fig.~\ref{fig:gpd} highlight the supervised outputs in {\color{red}red} and the regularized features in {\color{orange}orange}.

\section{Experiments}

We compare our approach with state-of-the-art methods \cite{dreamgaussian,ho2024sith,zhang2023sifu,zhuang2024idolinstantphotorealistic3d} for human reconstruction from a single image, demonstrating superior performance in both novel view synthesis and animation tasks (Tbl. \ref{tbl:comp_ahq_light} and Fig. \ref{fig:comp_ahq_light}). We then showcase the versatility of our framework through various applications like animation (Fig. \ref{fig:teaser}), editing and interpolation (Fig. \ref{fig:applications_light}). Finally, we conduct extensive ablation studies (Tbl. \ref{tbl:ablation}) to validate our design choices and analyze the impact of different components on the overall performance.

\subsection{Comparison with State-of-the-art}
\begin{figure*}[ht]
\begin{center}
\small
\setlength{\tabcolsep}{2pt}
\newcommand{\inputwidth}{1.4cm}
\begin{tabular}{c cccccc}
  \includegraphics[width=\inputwidth]{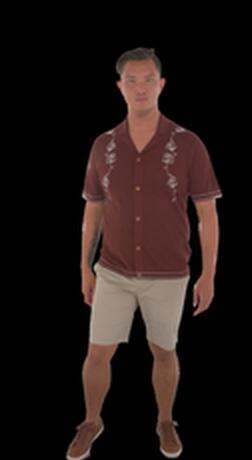}  
  & \includegraphics[width=\inputwidth]{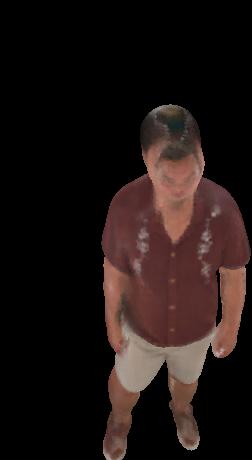}
  & \includegraphics[width=\inputwidth]{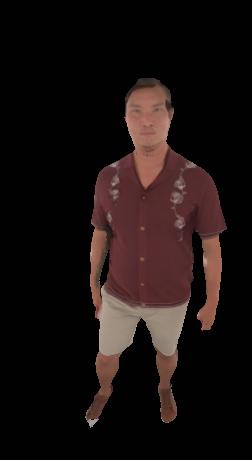}
  & \includegraphics[width=\inputwidth]{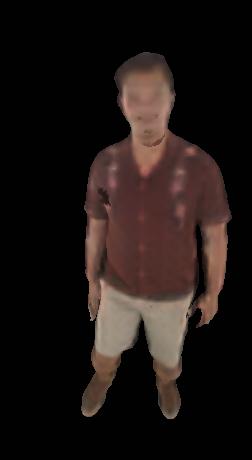}
  & \includegraphics[width=\inputwidth]{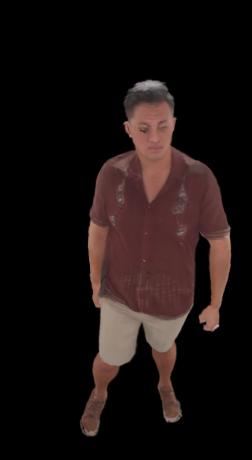}  
  & \includegraphics[width=\inputwidth]{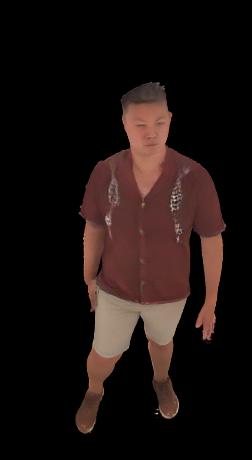} 
  & \includegraphics[width=\inputwidth]{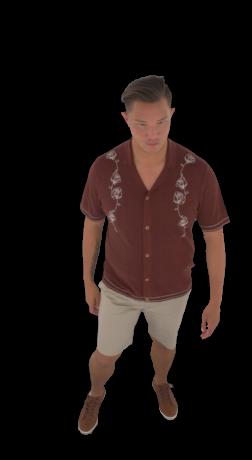}   \\
  \includegraphics[width=\inputwidth]{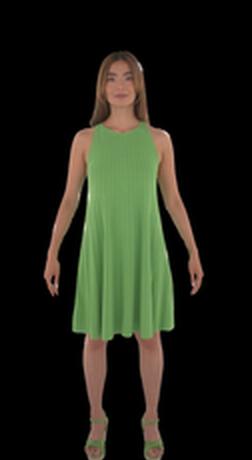}  
  & \includegraphics[width=\inputwidth]{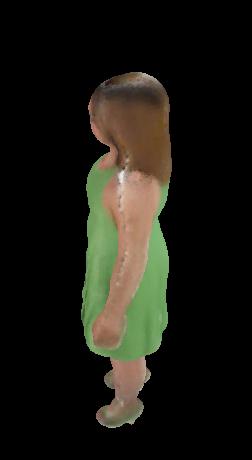}  
  & \includegraphics[width=\inputwidth]{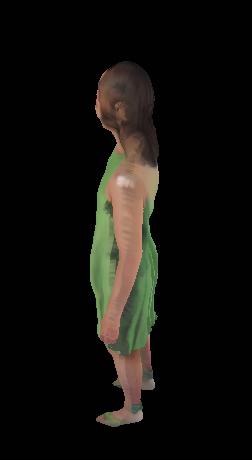}  
  & \includegraphics[width=\inputwidth]{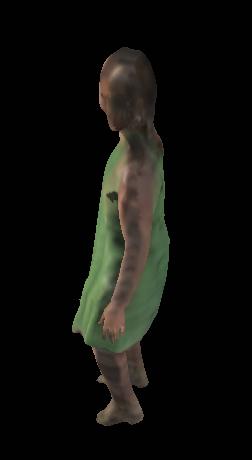}  
  & \includegraphics[width=\inputwidth]{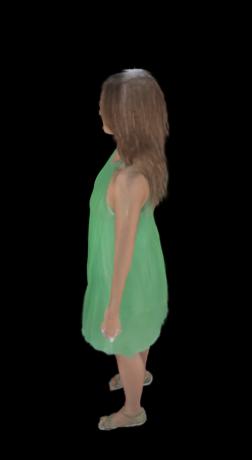}  
  & \includegraphics[width=\inputwidth]{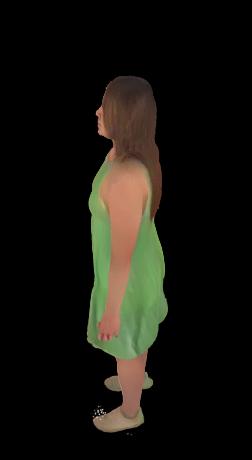} 
  & \includegraphics[width=\inputwidth]{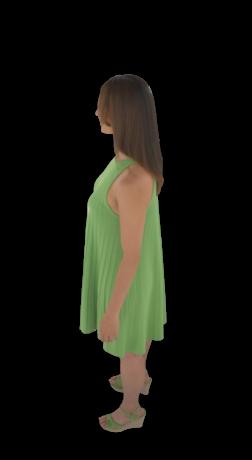}   \\
Input & DreamGaussian &  SiTH  & SIFU &  IDOL & \textbf{Ours} & GT
\end{tabular}
\end{center}
\caption{\label{fig:comp_ahq_light}Comparison for novel view synthesis with DreamGaussian \cite{dreamgaussian}, SiTH \cite{ho2024sith}, SIFU \cite{zhang2023sifu}, and IDOL \cite{li2023instant3d}.
Tbl. \ref{tbl:comp_ahq_light} lists metrics.
}
\end{figure*}

\begin{figure*}[ht]
\begin{center}
\small
\setlength{\tabcolsep}{2pt}
\newcommand{\inputwidth}{0.8cm}
\newcommand{\imagewidth}{8cm}
\begin{tabular}{c cccccc}
  \includegraphics[width=\inputwidth]{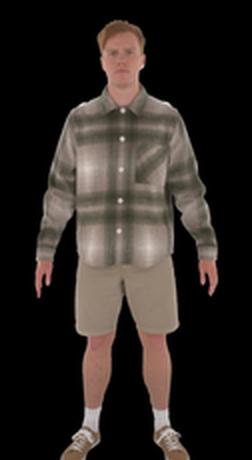}  &
  \includegraphics[width=\imagewidth]{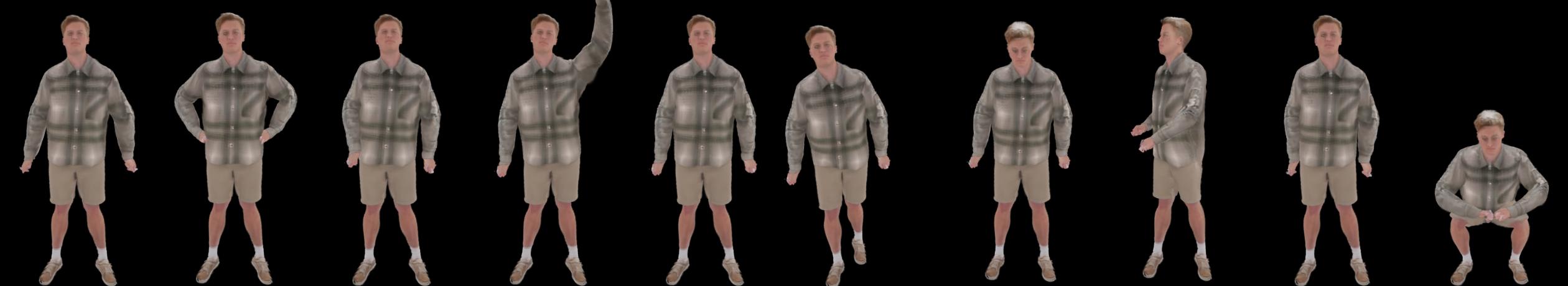} &
  \includegraphics[width=\imagewidth]{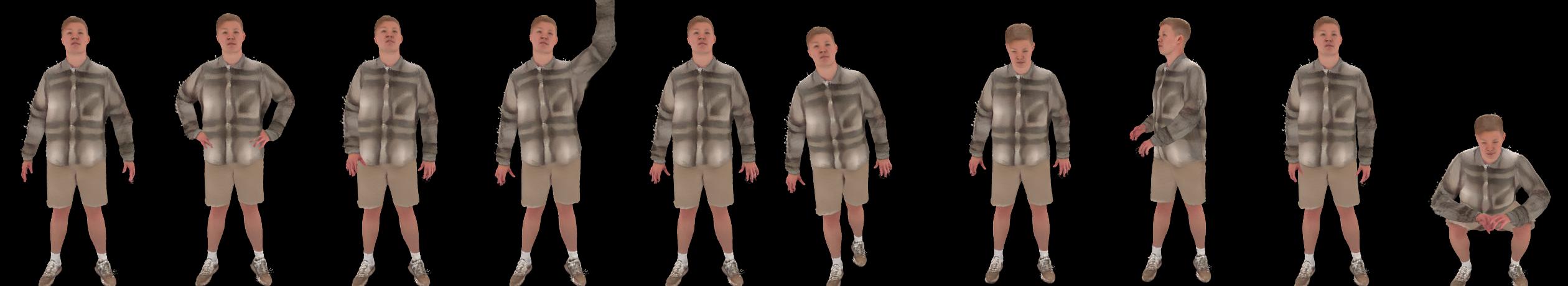} \\
  \includegraphics[width=\inputwidth]{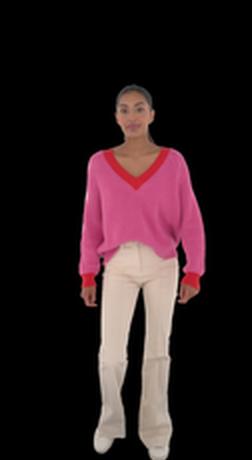}  &
  \includegraphics[width=\imagewidth]{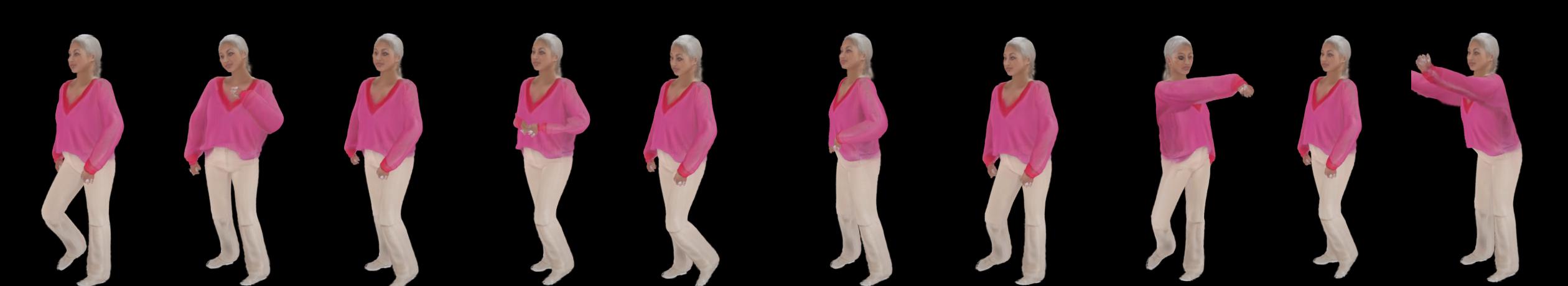} &
  \includegraphics[width=\imagewidth]{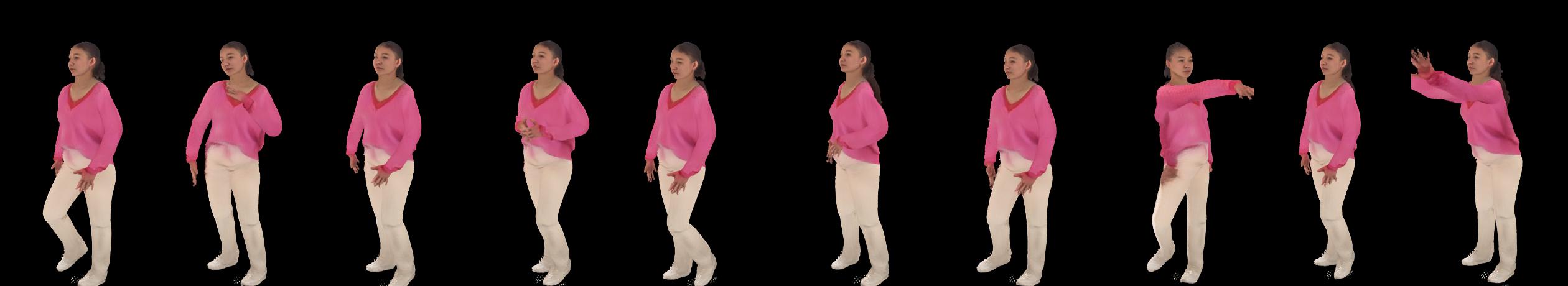} \\
  Input & IDOL~\cite{zhuang2024idolinstantphotorealistic3d} & Ours \\
\end{tabular}
\end{center}
\caption{\label{fig:animation_light}Comparison for animation. Our DLA framework enables detailed renderings for difficult poses, outperforming the state-of-the-art in one-shot animatable avatars in perceptual and photometric metrics. Please see the supp. mat. for more examples and metrics. \vspace{-1em}
}
\end{figure*}

\begin{figure*}[ht]
    \centering
  \includegraphics[width=\textwidth]{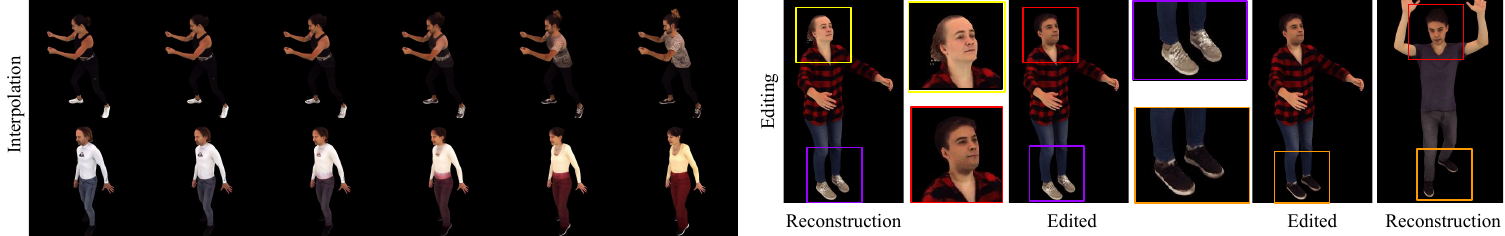}
    \caption{\label{fig:applications_light}Applications. Our structured latent code affords editing like face swapping and virtual try-on of shoes (right). In addition, we observe emerging capabilities like smooth interpolations between avatar latent codes (left). These examples are reconstructions using multi-view inputs from CustomHumans \cite{ho2023custom}. \vspace{-1em}}
\end{figure*}

We compare with several the state-of-the-art for one-shot human reconstruction \cite{zhuang2024idolinstantphotorealistic3d,ho2024sith,zhang2023sifu,dreamgaussian}. Tbl. \ref{tbl:comp_ahq_light} presents metrics for novel view synthesis. We measure perceptual quality (LPIPS) \cite{lpips}, structural similarity (SSIM), and photometric accuracy (PSNR).
Fig.~\ref{fig:animation_light} demonstrates animation capabilities while Fig.~\ref{fig:comp_ahq_light} shows visual results for novel view synthesis.

\begin{table}[ht]
\centering
\small 
\begin{tabular}{l  ccc }
\hline
\textbf{Method}
    & \textbf{LPIPS} $\downarrow$ & 
    \textbf{PSNR} $\uparrow$ &
    \textbf{SSIM} $\uparrow$
    \\
\hline
DreamGaussian \cite{dreamgaussian}
    & 0.1514 &  19.48 &    0.8837 \\
SiTH \cite{ho2024sith} & 0.1577 &  18.88 &    0.8764  \\
SIFU \cite{zhang2023sifu}& 0.1408 &   19.42 &    0.8823  \\
IDOL \cite{zhuang2024idolinstantphotorealistic3d}
    & 0.0696 & 24.48 & 0.9261 \\
\hline
\textbf{Ours} & \textbf{0.0580} & \textbf{25.58} & \textbf{0.9279} \\
\hline
\end{tabular}%
\caption{Quantitative results on ActorsHQ \cite{isik2023humanrf}. We compare novel view synthesis on the input image. Please see Fig. \ref{fig:comp_ahq_light} for visuals. The supplementary contains comparisons for novel pose synthesis and comparisons on the 4D-Dress dataset \cite{wang20244ddress}.
\label{tbl:comp_ahq_light}}
\vspace{-1em}
\end{table}


Our quantitative evaluation in Tbl.~\ref{tbl:comp_ahq_light} demonstrates superior performance in novel view synthesis. We compare our approach against state-of-the-art methods by taking a single image as input and rendering it from multiple unseen viewpoints, as visualized in Fig.~\ref{fig:comp_ahq_light}. The metrics reported in Tbl.~\ref{tbl:comp_ahq_light} confirm our method's effectiveness across all evaluation criteria. In the following, we discuss the results taking the limitations of the state-of-the-art into consideration.

DreamGaussian \cite{dreamgaussian} employs score distillation sampling \cite{poole2022dreamfusion} to produce static Gaussians that can be converted into a mesh with texture refinement, which does not account for view-dependent effects. SiTH \cite{ho2024sith} and SiFU \cite{zhang2023sifu} also reconstruct textured meshes using implicit functions, but require additional rigging (e.g., with Mixamo) for animation. IDOL \cite{zhuang2024idolinstantphotorealistic3d} offers fast inference but sacrifices quality, as it directly predicts a Gaussian map in the UV space. This presents two drawbacks: first, it attempts to solve two complex steps (Dream and Lift) simultaneously; second, their Gaussians are not conditioned on the target pose, ignoring pose-dependent effects. These limitations result in reduced reconstruction quality, as evidenced in Tbl.~\ref{tbl:comp_ahq_light} and Fig.~\ref{fig:comp_ahq_light}.

Furthermore, we showcase our animation capabilities on ActorsHQ \cite{isik2023humanrf} in Fig.~\ref{fig:animation_light} and on in-the-wild image \cite{fu2022styleganhuman} in Figures \ref{fig:teaser}. The results demonstrate high-quality pose transfer while preserving the avatar's geometric and appearance. Please see the supplementary material for a comparison for novel pose synthesis, comparisons on the 4D-Dress dataset \cite{wang20244ddress}, and experimental details regarding the comparisons.

\subsection{Applications}
\label{sec:applications}
The structured nature of our UV-space Gaussian Parameter Map, decoded from the avatar latent code $\mathbf{Z}$, enables a range of downstream applications. Beyond reconstruction and animation, it supports controllable editing and we observe emergent properties like smooth interpolations.

\noindent\textbf{Editing.}
Our structured latent representation $\mathbf{Z}$ is spatially aligned with the UV space of the human body, allowing semantic and part-aware manipulations. This alignment enables intuitive avatar editing by performing localized operations on $\mathbf{Z}$. For instance, we can swap specific regions between the latent codes of two different avatars. The decoder then faithfully reconstructs coherent and photorealistic avatars reflecting these edits. This facilitates flexible avatar customization and compositional synthesis (see Fig.~\ref{fig:applications_light}, bottom).

\noindent\textbf{Emerging Capabilities.}
Although our framework is not designed as a generative model, we observe that the learned latent space $\mathbf{Z}$ exhibits smooth structure similar to that of generative style spaces~\cite{stylegan,park2019SPADE,vqvae,buehler2021varitex,Rombach2022StableDiffusion}. As a result, interpolating between the latent codes of different avatars yields plausible and continuous transitions between identities.
Qualitative results are shown in Fig.~\ref{fig:applications_light} (top).

\noindent\textbf{Rendering Speed.}
The design of our Gaussian Parameter Decoder (Fig. \ref{fig:gpd}) allows real-time rendering while considering pose- and view-dependent effects. When rendering novel poses or views for a given subject, we cache the canonical Gaussian map $\mathbf{F}_c$ because it does not depend on the pose and viewpoint---only the offset map $\mathbf{F}_\Delta$ needs to be computed for each frame. This enables rendering frames of resolution  $512\times 512$ at 33 FPS (30 ms per frame) on a single NVIDIA RTX 5880.

\subsection{Ablations}
\label{sec:ablation}We report metrics for several ablation studies in Tbl. \ref{tbl:ablation}. The first set of ablations examines techniques for mapping pose-space Gaussians from an unstructured cloud to a structured UV space (A). In ablation A.i, we use the SMPL-X body model to directly project pixel-aligned Gaussians to the UV space, similar to RGB texture unprojection when reconstructing textured meshes from multi-view inputs. This approach lacks learning capabilities and fails to account for shapes beyond the body model (like clothing), resulting in a significant drop in perceptual quality (27\% decrease in LPIPS). We also experiment with learning the query (A.ii) instead of using the vertex position map ($\mathbf{X}$ in Fig. \ref{fig:encoder}) and found it yields worse results.  We conclude that the vertex position map introduces valuable prior knowledge from the body model, which is missing in the learned query.
The second ablation (B) finds a positive impact of view- and pose-conditional inputs in the Gaussian Parameter Decoder (Fig. \ref{fig:gpd}). 
The third set of ablations (C) considers different inputs to our full model. 
For inference, our encoder takes the ground-truth frontal input image $\mathbf{I}_1$, three synthesized images $\{ \mathbf{I}_2^n, ..., \mathbf{I}_V^n\}$, and estimated SMPL-X parameters as input. We study the effect of noisy SMPL-X parameters in the transformer encoder (C.i), which is particularly relevant for in-the-wild settings where fitting the body model might be challenging.  Replacing the ground-truth input image $\mathbf{I}_1$ with the reconstruction from the diffusion model slightly reduces performance (C.ii). Conversely, we simulate the potential of next-generation video diffusion models by feeding only ground-truth novel views as input, which leads to a substantial performance gain (C.iii), indicating room for further improvement by just replacing the video diffusion model with an improved version, thanks to our our modular design. We provide more detailed ablations in the supp. mat.

\begin{table}[ht]
\footnotesize 
\begin{center}
\begin{tabular}{l ccc }
\hline
    & \textbf{LPIPS} $\downarrow$ & 
    \textbf{PSNR} $\uparrow$ &
    \textbf{SSIM} $\uparrow$
    \\
\hline
\textbf{A. UV mapping} \\
i) Mesh unprojection (no learning) & 0.0790 & 23.66 & 0.9209 \\
ii) Learned Query & 0.0652 & 24.67 & 0.9254 \\
\hline
\textbf{B. Animate \& render}\\
No conditioning & 0.0644 & 24.86 & 0.9248 \\
\hline
\textbf{C. Inputs} \\
i) Noisy SMPL-X inputs & 0.0660 & 24.77 & 0.9240 \\
ii) 4 synthetic inputs & 0.0657 & 24.78 & 0.9237 \\
iii) 4 GT inputs & 0.0613 & 25.71 & 0.9310 \\
\hline
\textbf{Ours} & 0.0624 & 25.09 & 0.9254 \\
\hline
\end{tabular}
\caption{Ablation studies (Sec. \ref{sec:ablation}).\label{tbl:ablation}}
\end{center}
\vspace{-4em}
\end{table}

\section{Conclusion}

We introduced a novel framework for high-quality 3D human avatar reconstruction from a single image. Our approach, Dream, Lift, Animate, leverages a multiview diffusion model to \textit{dream} plausible unseen viewpoints, which are then \textit{lifted} into an unstructured 3D Gaussian representation via a large feed-forward reconstruction model. A key contribution of our work is a transformer-based encoder that maps these unstructured Gaussians into a structured UV-space representation. This structured form enables realistic animation, fine-grained control, and intuitive editing, while also exhibiting emergent properties such as smooth identity interpolation. Through extensive experiments, we demonstrate that our method achieves state-of-the-art performance in both perceptual quality and photometric accuracy across multiple benchmarks.

Our method has several limitations that warrant discussion. Close body parts may cause color leakage between adjacent regions (e.g., hand skin color affecting nearby clothing). Moreover, while our method can handle slight inconsistencies in the generated multiview images, it cannot handle significant inconsistencies. Finally, reconstructed avatars may exhibit inconsistencies in identity, primarily in facial features. This stems from training dataset biases (THuman2.0, CustomHumans, and 4D-Dress) and the low resolution of face regions in the input. Training our method on in-the-wild monocular images is an interesting future work.
{
    \small
    \bibliographystyle{ieeenat_fullname}
    \bibliography{main}
}

\appendix
\clearpage

This supplementary describes the model architecture and experimental setting in more detail in Sec. \ref{suppsec:details}, provides supplementary comparisons and ablations in Sec. \ref{suppsec:results}, and discusses the potential societal impact of this work in Sec. \ref{suppsec:impact}.

\section{Experimental and Model Details}
\label{suppsec:details}
\paragraph{Training Datasets}
\label{suppsec:dataset}
We train our autoencoder on CustomHumans~\cite{ho2023custom}, 4D-Dress~\cite{wang20244ddress}, and THuman2.0~\cite{tao2021function4d}. 
We randomly select 597 scans from CustomHumans, 
419 sequences from 4D-Dress,
and 519 scans from THuman. We render the scanned meshes from 20 views for CustomHumans, from 54 views for THuman2.0, and from 24 views for 4D-Dress.
Note that these datasets are limited in terms of diversity. Customhumans has 81, 4D-Dress 32, and THuman2.0 500 subjects.

\paragraph{Evaluation}
For evaluation, we compare our method with state-of-the-art approaches on ActorsHQ~\cite{isik2023humanrf}, holdout subjects from 4D-Dress~\cite{wang20244ddress}, and demonstrate in-the-wild qualitative results on SHHQ \cite{fu2022styleganhuman}. On ActorsHQ, we evaluate across all available sequences and subsample by a factor of 100 to obtain approximately 20-25 frames per sequence. We render from a carefully selected subset of 14 camera viewpoints that capture the full body from diverse angles. For 4D-Dress, we select 2 sequences from each of 6 unseen subjects, yielding 12 evaluation sequences. We render 24 views for each scan with perspective cameras from a distance of 2.4m. The 4D-Dress sequences are subsampled by a factor of 10, providing 15-20 diverse frames per sequence.
To show in-the-wild generalization of our approach, we run inference on images from SHHQ \cite{fu2022styleganhuman}.
All evaluations are conducted at resolution $512\times 512$ with a black background.

We report metrics for novel view synthesis, novel pose synthesis, and both combined. For novel view synthesis, we input a frontal view of the first frame for each sequence and compute metrics on novel views for this frame. For novel pose synthesis, we render the full sequences while only considering the input camera. For novel view \& pose, we compute metrics on multiple camera views for the full sequence. 

To allow reproducibility on the public benchmark, we report quantitative numbers on models that are trained on public datasets. To improve our generalization on in-the-wild images, we also trained a model using synthetic data generated using the multiview diffusion model~\cite{wang2024unianimate}. 

\subsection{Implementation Details}
\label{suppsec:implementationdetails}
The model for generating pseudo-ground-truth novel views is initialized with pretrained weights from LGM \cite{tang2025lgm}. We finetune this model during the first 10 epochs of training. For consistency, we canonicalize all SMPL-X meshes by centering the pelvis at the origin. Our training follows a progressive approach: 10 epochs at $256 \times 256$ resolution followed by 10 epochs at $512 \times 512$ resolution. We use a batch size of 32, with each sample comprising 3 randomly selected output views. For the transformer component of our encoder, we adapt and extend the architecture from 3DShape2VecSet~\cite{zhang3dshape2vecset}. During the finetuning stage, we generate pseudo-GT views using UniAnimate~\cite{wang2024unianimate}.
To create clean silhouettes for the pseudo-GT images, we apply the rembg~\cite{Gatis_rembg_2025} background removal tool to the outputs.

\paragraph{Architectural Details}

The pose-space reconstruction model receives four images at resolution $256 \times 256$. The \emph{lift} step produces four pose-space Gaussian maps of resolution $128 \times 128$, resulting in $65,536$ Gaussians. The filtering step removes Gaussians with low opacity. Typically, about 40-50\% of the initial Gaussians remain as input to the encoder $\mathcal{F}$. The latent has 256 channels and spatial dimensions $64\times 64$. The output of the Gaussian Parameter decoder has shape $14\times 256\times256$. We uniformly sample 65,536 Gaussians in the UV space and 65,536 on the mesh surface, resulting in $131,072$ Gaussians in total.

The number of parameters is 419 M for the pose-space reconstruction model, 19M for the feature projections and transformer encoder $\mathcal{F}$, and 12 M for the Gaussian Parameter Decoder. The patch discriminator \cite{isola2017image} has 7 M parameters.

\paragraph{Training Details}
We only apply the loss for the unstructured Gaussians $\mathcal{L}_\text{UG}$ and the L1 reconstruction loss $\mathcal{L}_\text{L1}$ for low-resolution training (during the first 10 epochs). We optimize with Adam with momentum $\beta=(0.5, 0.9)$ and constant learning rate $0.0001$ for the generator and $0.000001$ for the discriminator.

\paragraph{Inference-time Multi-view Generation}
As described in the main paper, we synthesize novel views using a ControlNet-guided variant~\cite{wang2024unianimate} of a video diffusion model~\cite{wan2025}. Specifically, we first estimate the SMPL-X parameters from the input image \cite{he2024magicman}. We then render 2D skeletal poses of the predicted mesh from virtual cameras placed around a 360-degree azimuth with a 0-degree elevation angle. These projected poses serve as control signals to guide the diffusion model in generating photorealistic images from novel viewpoints. We render a video of a 360-degree azimuth rotation in 80 frames and pick the frames 20, 40, 60, and 80 as inputs to our model. 
For in-the-wild inputs, where major self-contact is common (e.g., hands hidden in the pocket), we condition on an A-pose skeleton. 
As in the original video diffusion model \cite{wang2024unianimate}, we use DDIM sampling \cite{ddim} with 50 steps.

\paragraph{Compute Resources}
This project was developed on a SLURM cluster. The complete training process, including logging and regular validation, requires 24 hours on 8 A100 80GB GPUs. Training with a batch size of 32, as specified in Sec. \ref{suppsec:implementationdetails}, uses 52 GB of VRAM. Inference only requires about 4 GB of VRAM. The full project consumed more resources for initial experiments and ablations. In addition, the project involved GPU and CPU resources for preprocessing and rendering scans from multiple datasets (specified in Sec. \ref{suppsec:dataset}).

\subsection{Related Works Details}
\paragraph{IDOL}
We input the image at $1024\times 1024$ resolution and render with a black background. All other hyperparameters are left as provided in the GitHub repository\footnote{https://github.com/yiyuzhuang/IDOL}. For the comparison on 4D-Dress, we observe a degraded performance, which we attribute to a misalignment in the body model. The IDOL code base only supports a SMPL-X body model with a neutral gender and provides precomputed files. 4D-Dress, however, uses the female and male body models. We reached out to the IDOL authors and asked for the respective cache files for the (fe-)male model via two different channels, but did not receive an answer. Hence, we ran inference on the neutral body model. ActorsHQ \cite{isik2023humanrf} provides SMPL-X parameters for the neutral gender, which are compatible with the IDOL codebase. Note that, in addition to publicly available 3D scans, IDOL also uses 100K synthetic multiview images for training.  Whereas our models are only trained on publicly available datasets with far fewer training samples, yet our method significantly outperforms IDOL, demonstrating room for further improvement with more training data. Note that IDOL trains on 100K synthetic images, whereas our model is only trained on public data.

\paragraph{SiTH}
We obtain the code and pretrained model from the official repository\footnote{https://github.com/SiTH-Diffusion/SiTH} and run their code without modifications. To compute the metrics, we compute an alignment between the output SMPL-X body with the ground-truth SMLP-X body from the dataset via iterative closest point. This alignment is applied to the output mesh and rendered to an image.

\paragraph{SIFU}
As for SiTH, we obtain the code and pretrained model from the official repository\footnote{https://github.com/River-Zhang/SIFU}. The default code does output colored meshes, hence, we uncomment a line in the script to set the vertex colors. Besides this minor change, the code is run as provided. To compute the metrics, we compute an alignment between the output SMPL-X body with the ground-truth SMLP-X body from the dataset via iterative closest point. This alignment is applied to the output mesh and rendered to an image.

\paragraph{DreamGaussian} Given a single-view testing image, we follow the DreamGaussian pipeline\footnote{https://github.com/dreamgaussian/dreamgaussian} to generate a 3D avatar. First, we optimize 3D Gaussians using a combination of reconstruction loss and SDS (Score Distillation Sampling) loss. Subsequently, we extract a coarse mesh from these optimized Gaussians and apply further refinement using similar objective functions. Throughout this process, we leverage Stable-Zero123 as our diffusion prior to ensure best performance. For quantitative evaluation, we render the final mesh from the corresponding camera viewpoint and calculate the metrics outlined in Section 4.1 of our paper.

\section{Supplementary Results}
This section provides supplementary comparisons and ablations.
\label{suppsec:results}
\begin{figure*}[ht]
    \centering
\small
\setlength{\tabcolsep}{2pt}
\newcommand{\inputheight}{2.5cm}
\newcommand{\imagewidth}{1.5cm}
\begin{tabular}{c c}
  \includegraphics[height=\inputheight]{fig/teaser/image_000021_input.jpg}  &
  \includegraphics[height=\inputheight]{fig/teaser/image_000021_output.jpg} \\
  \includegraphics[height=\inputheight]{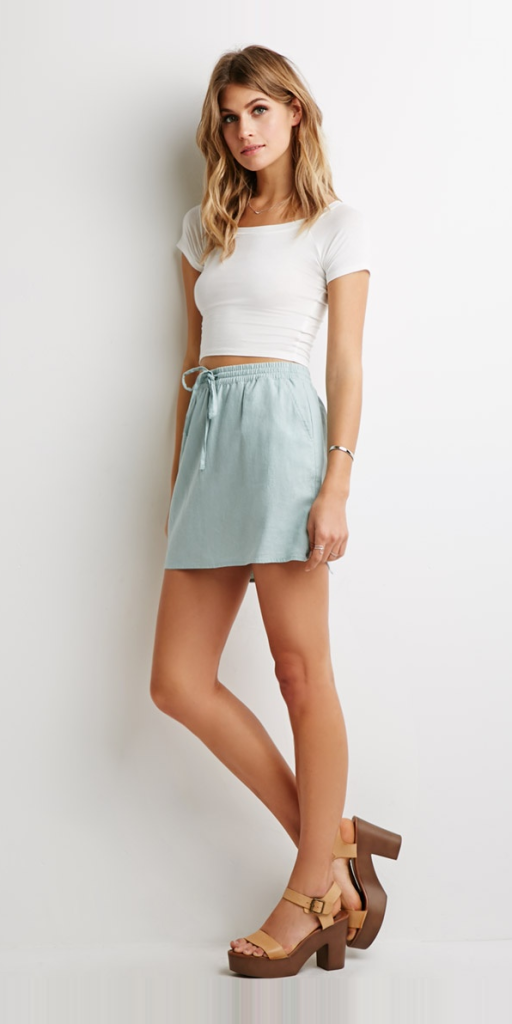}  &
  \includegraphics[height=\inputheight]{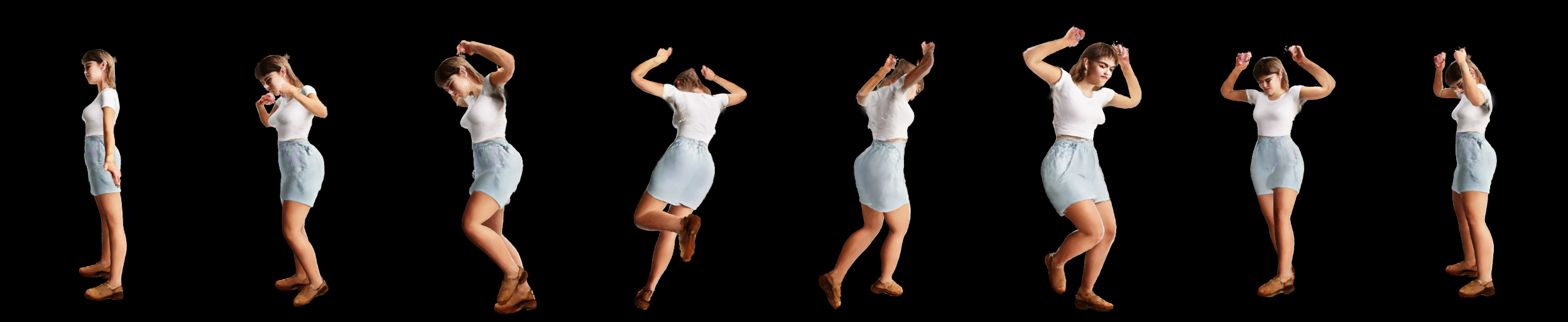} \\
  \includegraphics[height=\inputheight]{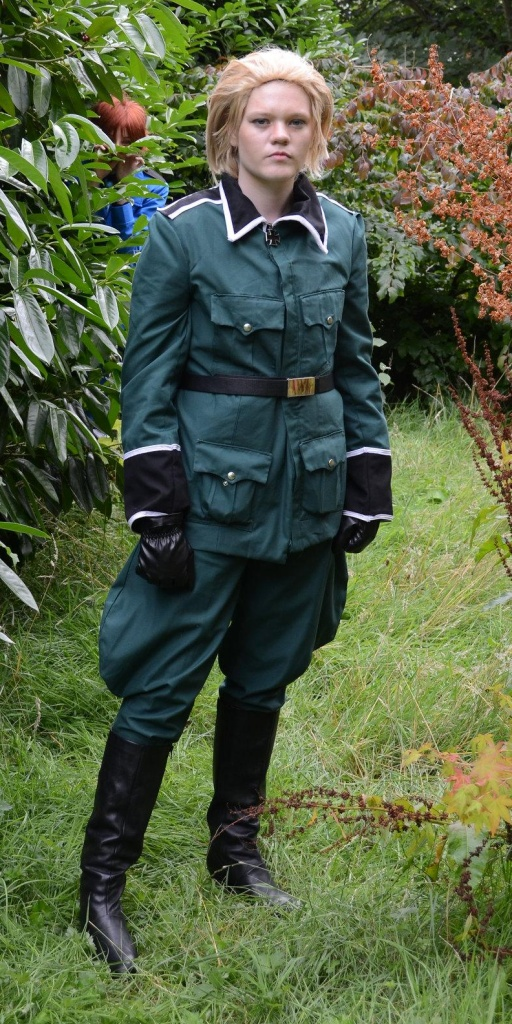}  &
  \includegraphics[height=\inputheight]{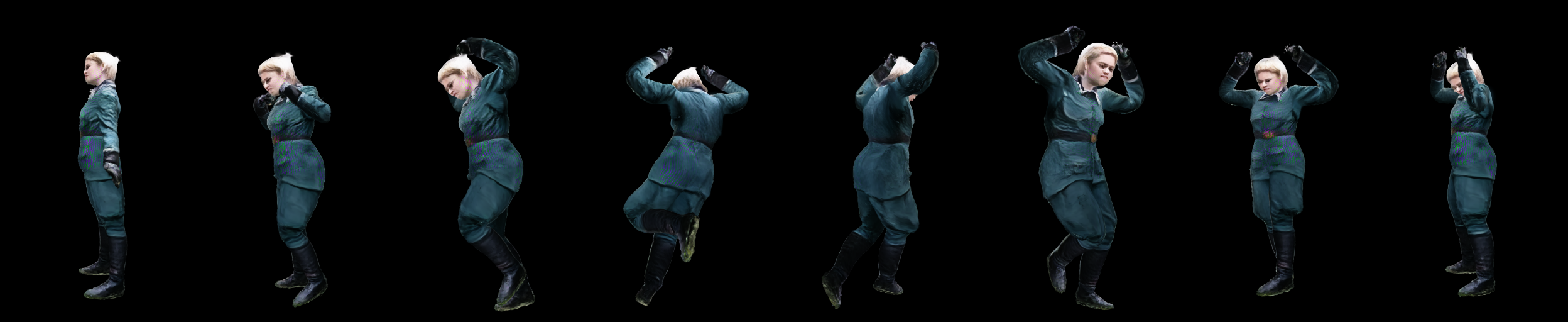} \\
  Input & Novel Views and Poses \\
\end{tabular}
\caption{\label{fig:shhq} We animate single input images from SHHQ \cite{fu2022styleganhuman} and render novel views while changing the body pose, demonstrating the robustness of our method to in-the-wild inputs. See the teaser figure in the main paper for more examples.
}
\end{figure*}

\begin{figure*}[ht]
\begin{center}
\small
\setlength{\tabcolsep}{2pt}
\newcommand{\inputwidth}{1.4cm}
\begin{tabular}{c cccccc}
  \includegraphics[width=\inputwidth]{fig/comp_ahq/Actor02_Sequence2_000000_7_input.jpg}  
  & \includegraphics[width=\inputwidth]{fig/comp_ahq/Actor02_Sequence2_000000_7_dreamgaussian.jpg}
  & \includegraphics[width=\inputwidth]{fig/comp_ahq/Actor02_Sequence2_000000_7_sith.jpg}
  & \includegraphics[width=\inputwidth]{fig/comp_ahq/Actor02_Sequence2_000000_7_sifu.jpg}
  & \includegraphics[width=\inputwidth]{fig/comp_ahq/Actor02_Sequence2_000000_7_idol.jpg}  
  & \includegraphics[width=\inputwidth]{fig/comp_ahq/Actor02_Sequence2_000000_7_ours.jpg} 
  & \includegraphics[width=\inputwidth]{fig/comp_ahq/Actor02_Sequence2_000000_7_gt.jpg}   \\
  \includegraphics[width=\inputwidth]{fig/comp_ahq/Actor01_Sequence1_000000_94_input.jpg}  
  & \includegraphics[width=\inputwidth]{fig/comp_ahq/Actor01_Sequence1_000000_94_dreamgaussian.jpg}  
  & \includegraphics[width=\inputwidth]{fig/comp_ahq/Actor01_Sequence1_000000_94_sith.jpg}  
  & \includegraphics[width=\inputwidth]{fig/comp_ahq/Actor01_Sequence1_000000_94_sifu.jpg}  
  & \includegraphics[width=\inputwidth]{fig/comp_ahq/Actor01_Sequence1_000000_94_idol.jpg}  
  & \includegraphics[width=\inputwidth]{fig/comp_ahq/Actor01_Sequence1_000000_94_ours.jpg} 
  & \includegraphics[width=\inputwidth]{fig/comp_ahq/Actor01_Sequence1_000000_94_gt.jpg}   \\
  \includegraphics[width=\inputwidth]{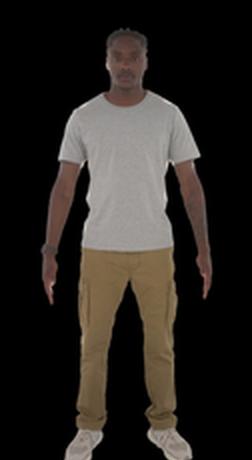}  
  & \includegraphics[width=\inputwidth]{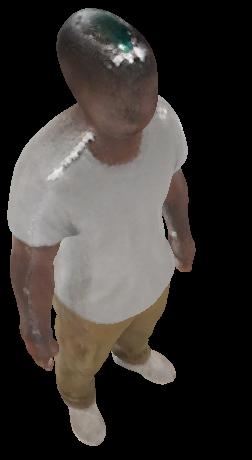}  
  & \includegraphics[width=\inputwidth]{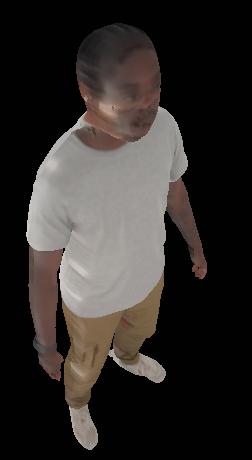}  
  & \includegraphics[width=\inputwidth]{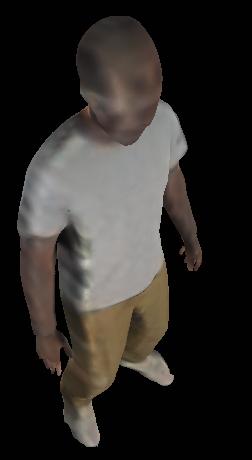}  
  & \includegraphics[width=\inputwidth]{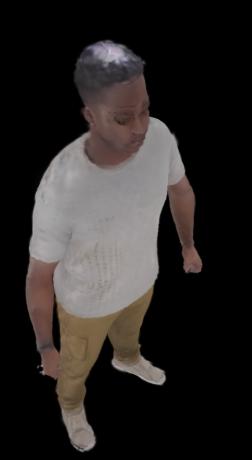}  
  & \includegraphics[width=\inputwidth]{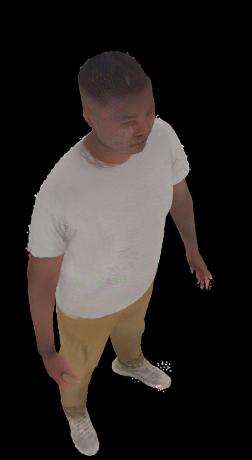} 
  & \includegraphics[width=\inputwidth]{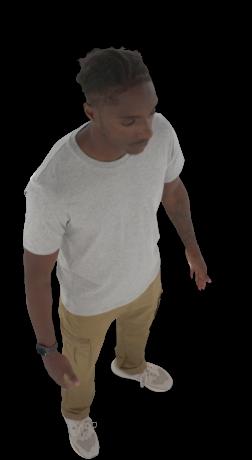}   \\
  \includegraphics[width=\inputwidth]{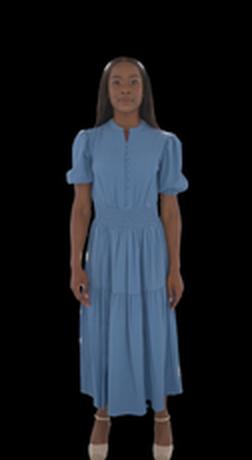}  
  & \includegraphics[width=\inputwidth]{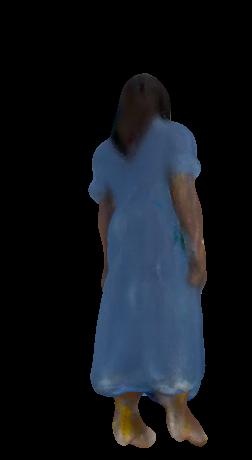}  
  & \includegraphics[width=\inputwidth]{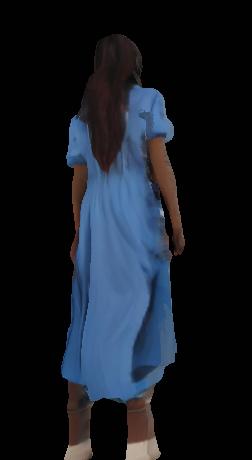}  
  & \includegraphics[width=\inputwidth]{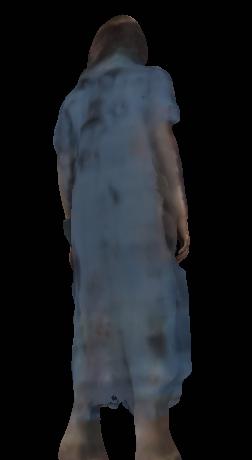}  
  & \includegraphics[width=\inputwidth]{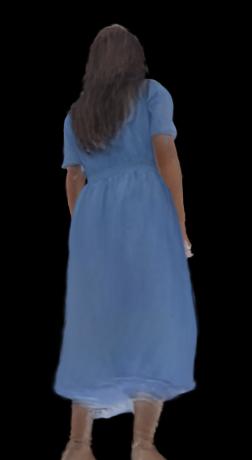}  
  & \includegraphics[width=\inputwidth]{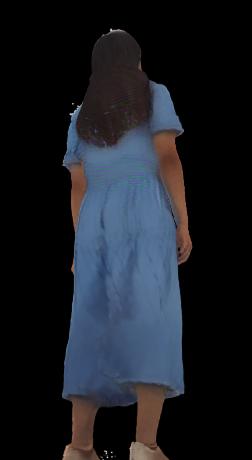} 
  & \includegraphics[width=\inputwidth]{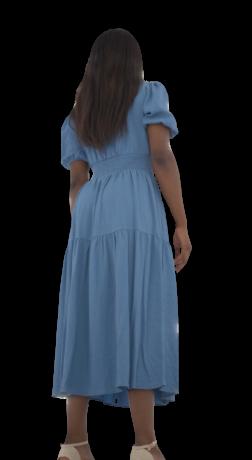}   \\
Input & DreamGaussian &  SiTH  & SIFU &  IDOL & \textbf{Ours} & GT
\end{tabular}
\end{center}
\caption{\label{fig:comp_ahq}Comparison for novel view synthesis with DreamGaussian \cite{dreamgaussian}, SiTH \cite{ho2024sith}, SIFU \cite{zhang2023sifu}, and IDOL \cite{li2023instant3d}.
Please see Table \ref{tbl:comp_ahq} for metrics.\vspace{-1.5em}
}
\end{figure*}
\begin{figure*}[ht]
\begin{center}
\small
\setlength{\tabcolsep}{2pt}
\newcommand{\inputwidth}{0.85cm}
\newcommand{\imagewidth}{8.5cm}
\begin{tabular}{c cccccc}
  \includegraphics[width=\inputwidth]{fig/comp_ahq_anim/Actor08_Sequence1_anim_126_input.jpg}  &
  \includegraphics[width=\imagewidth]{fig/comp_ahq_anim/Actor08_Sequence1_anim_126_idol.jpg} &
  \includegraphics[width=\imagewidth]{fig/comp_ahq_anim/Actor08_Sequence1_anim_126_ours.jpg} \\
  \includegraphics[width=\inputwidth]{fig/comp_ahq_anim/Actor04_Sequence2_anim_125_input.jpg}  &
  \includegraphics[width=\imagewidth]{fig/comp_ahq_anim/Actor04_Sequence2_anim_125_idol.jpg} &
  \includegraphics[width=\imagewidth]{fig/comp_ahq_anim/Actor04_Sequence2_anim_125_ours.jpg} \\
  \includegraphics[width=\inputwidth]{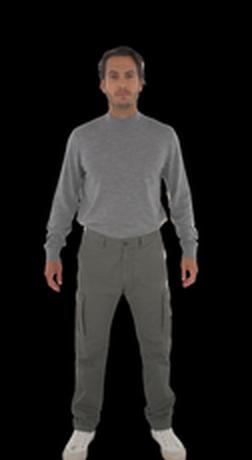}  &
  \includegraphics[width=\imagewidth]{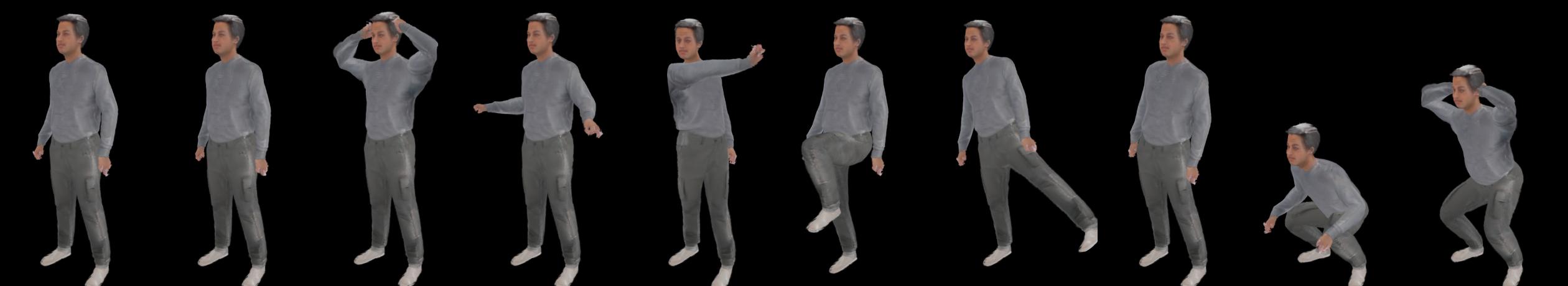} &
  \includegraphics[width=\imagewidth]{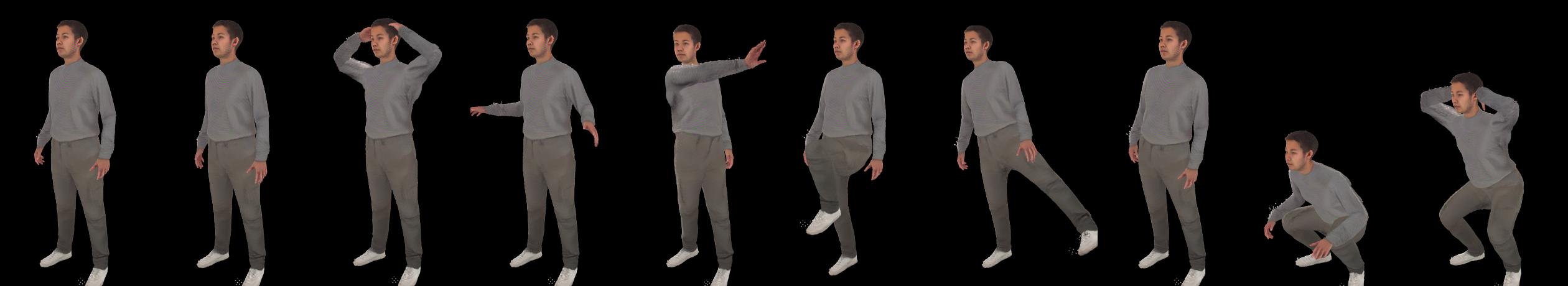} \\
  \includegraphics[width=\inputwidth]{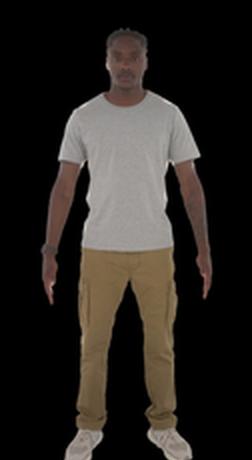}  &
  \includegraphics[width=\imagewidth]{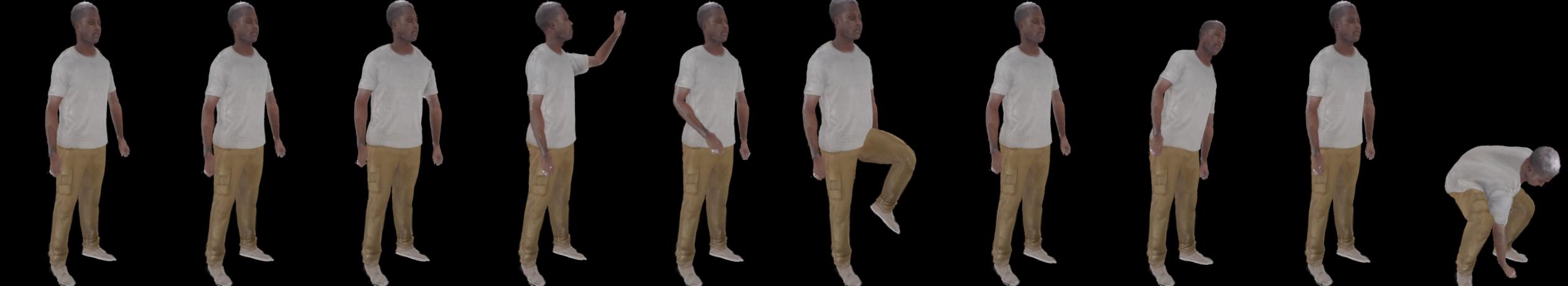} &
  \includegraphics[width=\imagewidth]{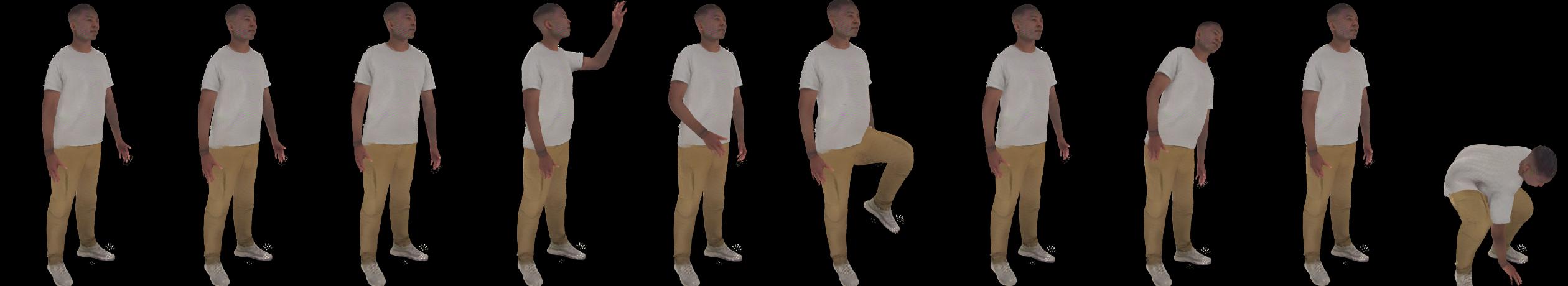} \\
  Input & IDOL~\cite{zhuang2024idolinstantphotorealistic3d} & Ours \\
\end{tabular}
\end{center}
\caption{\label{fig:animation}Comparison for animation with IDOL \cite{zhuang2024idolinstantphotorealistic3d}. Our Dream, Lift, Animate framework enables detailed renderings for difficult poses, outperforming the state-of-the-art in one-shot animatable avatars in perceptual and photometric metrics. Please see Table \ref{tbl:comp_ahq} for metrics.
}
\end{figure*}

\begin{table*}[ht]
\small 
\resizebox{\textwidth}{!}{%
\begin{tabular}{l  ccc  ccc ccc }
\hline
    & \multicolumn{3}{c}{\textbf{Novel view}} 
    & \multicolumn{3}{c}{\textbf{Novel pose}}
    & \multicolumn{3}{c}{\textbf{Novel view \& pose}} \\
 \\
\textbf{Method}
    & \textbf{LPIPS} $\downarrow$ & 
    \textbf{PSNR} $\uparrow$ &
    \textbf{SSIM} $\uparrow$
    & \textbf{LPIPS} $\downarrow$ & 
    \textbf{PSNR} $\uparrow$ &
    \textbf{SSIM} $\uparrow$
    & \textbf{LPIPS} $\downarrow$ & 
    \textbf{PSNR} $\uparrow$ &
    \textbf{SSIM} $\uparrow$
    \\
\hline
DreamGaussian \cite{dreamgaussian}
    & 0.1514 &  19.48 &    0.8837 & - & - & - & - & - & - \\
SiTH \cite{ho2024sith} & 0.1577 &  18.88 &    0.8764 & - & - & - & - & - & - \\
SIFU \cite{zhang2023sifu}& 0.1408 &   19.42 &    0.8823 & - & - & - & - & - & - \\
IDOL \cite{zhuang2024idolinstantphotorealistic3d}
    & 0.0696 & 24.48 & 0.9261 & 0.0940 & 22.80 & 0.9194 & 0.1022 & 22.26 & 0.9136 \\
\hline
\textbf{Ours} & \textbf{0.0580} & \textbf{25.58} & \textbf{0.9279} & \textbf{0.0471} &\textbf{26.41} & \textbf{0.9351} & \textbf{0.0624} & \textbf{25.09} & \textbf{0.9254} \\
\hline
\end{tabular}%
}
\caption{Quantitative results on ActorsHQ \cite{isik2023humanrf}. We compare extend the comparison from the main paper on novel view synthesis with novel poses using the input camera, and novel views and novel poses jointly. Only IDOL is readily animatable---the other related works would require postprocessing like rigging the reconstructed mesh. Please see Fig. \ref{fig:comp_ahq} for visuals.
\label{tbl:comp_ahq}}
\end{table*}

\begin{table*}[ht]
\small 

\begin{center}
\begin{tabular}{l  ccc  ccc ccc }
\hline

    & \multicolumn{3}{c}{\textbf{Novel view}} 
    & \multicolumn{3}{c}{\textbf{Novel pose}}
    & \multicolumn{3}{c}{\textbf{Novel view \& pose}} \\
 \\
\textbf{Method}
    & \textbf{LPIPS} $\downarrow$ & 
    \textbf{PSNR} $\uparrow$ &
    \textbf{SSIM} $\uparrow$
    & \textbf{LPIPS} $\downarrow$ & 
    \textbf{PSNR} $\uparrow$ &
    \textbf{SSIM} $\uparrow$
    & \textbf{LPIPS} $\downarrow$ & 
    \textbf{PSNR} $\uparrow$ &
    \textbf{SSIM} $\uparrow$
    \\
\hline
IDOL & 0.0904 & 23.04 & 0.9226 & 0.0956 & 22.54 & 0.9135 & 0.0993 & 21.99 & 0.9135 \\
\hline
\textbf{Ours} & \textbf{0.0594} & \textbf{24.95} & \textbf{0.9294} & \textbf{0.0741} & \textbf{23.25} & \textbf{0.9155} & \textbf{0.0775} & \textbf{23.25} & \textbf{0.9177} \\
\hline
\end{tabular}\caption{Quantitative results on 4D-Dress \cite{wang20244ddress}. We compare novel view synthesis on the input image, novel poses using the input camera, and novel views and novel poses. Please see Fig. \ref{fig:comp_4d} for visuals.
\label{tbl:comp_4d}}
\end{center}
\end{table*}
\begin{figure*}[ht]
\begin{center}
\small
\setlength{\tabcolsep}{2pt}
\newcommand{\inputwidth}{2cm}
\begin{tabular}{c ccc}
  \includegraphics[width=\inputwidth]{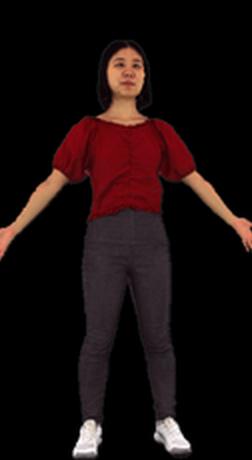}  
  & \includegraphics[width=\inputwidth]{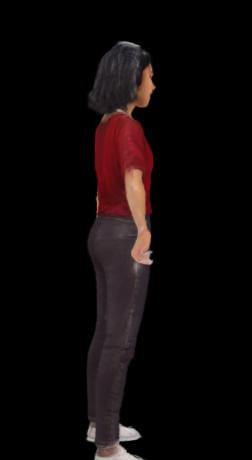}  
  & \includegraphics[width=\inputwidth]{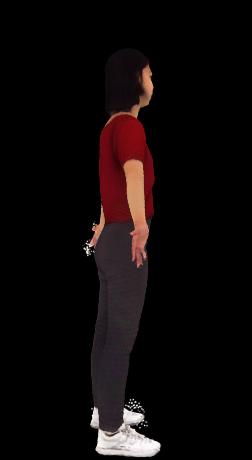} 
  & \includegraphics[width=\inputwidth]{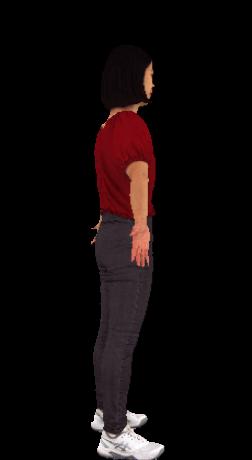}   \\
  \includegraphics[width=\inputwidth]{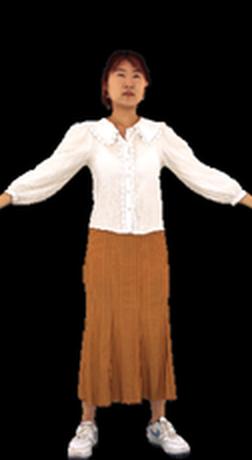}  
  & \includegraphics[width=\inputwidth]{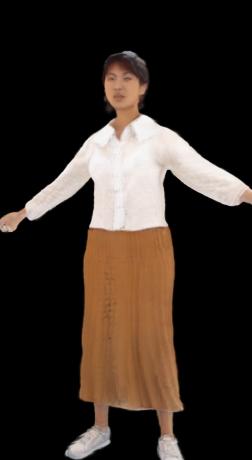}  
  & \includegraphics[width=\inputwidth]{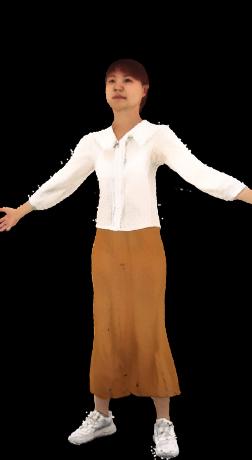} 
  & \includegraphics[width=\inputwidth]{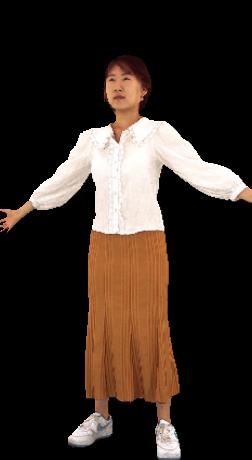}   \\
  \includegraphics[width=\inputwidth]{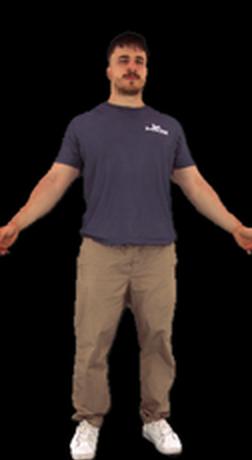}  
  & \includegraphics[width=\inputwidth]{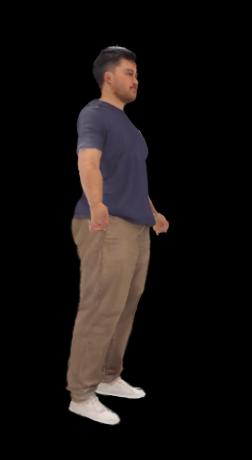}  
  & \includegraphics[width=\inputwidth]{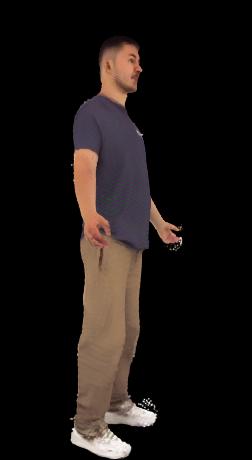} 
  & \includegraphics[width=\inputwidth]{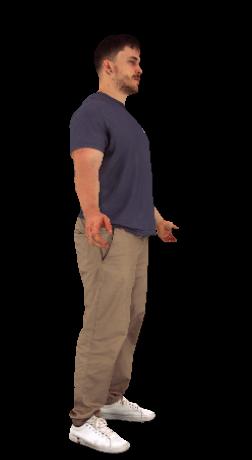}   \\
  \includegraphics[width=\inputwidth]{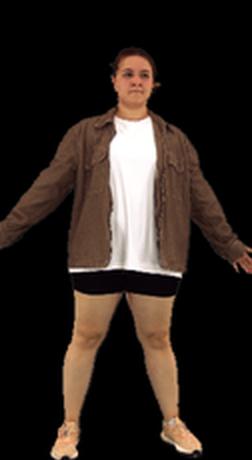}  
  & \includegraphics[width=\inputwidth]{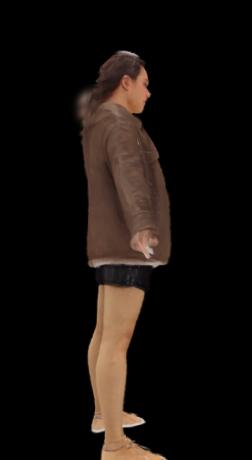}  
  & \includegraphics[width=\inputwidth]{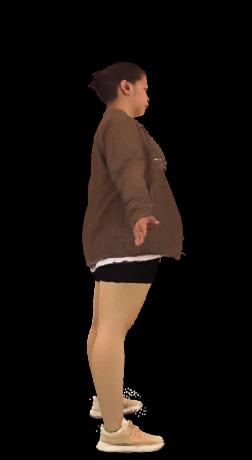} 
  & \includegraphics[width=\inputwidth]{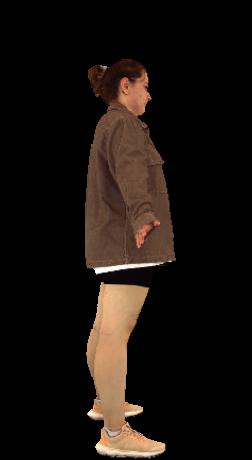}   \\
  \includegraphics[width=\inputwidth]{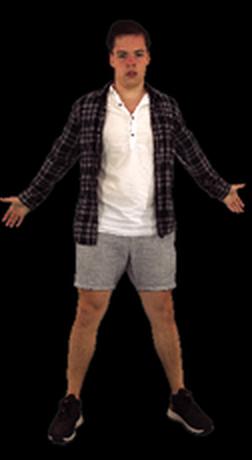}  
  & \includegraphics[width=\inputwidth]{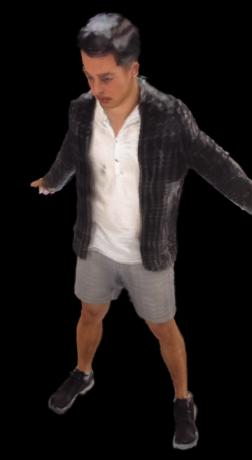}  
  & \includegraphics[width=\inputwidth]{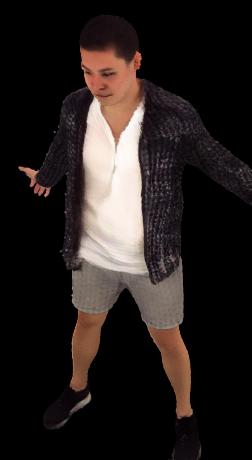} 
  & \includegraphics[width=\inputwidth]{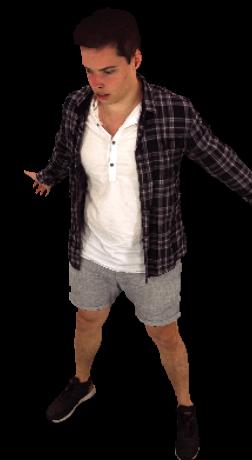}   \\
Input &  IDOL & \textbf{Ours} & GT
\end{tabular}
\end{center}
\vspace{-3mm}
\caption{\label{fig:comp_4d}Comparison for novel view synthesis on 4D-Dress \cite{wang20244ddress} with IDOL \cite{li2023instant3d}.
Please see Table \ref{tbl:comp_4d} for metrics.
}
\end{figure*}
\begin{figure*}[ht]
\begin{center}
\small
\setlength{\tabcolsep}{2pt}
\newcommand{\inputwidth}{0.65cm}
\newcommand{\imagewidth}{6.5cm}
\begin{tabular}{c cccccc}
  \includegraphics[width=\inputwidth]{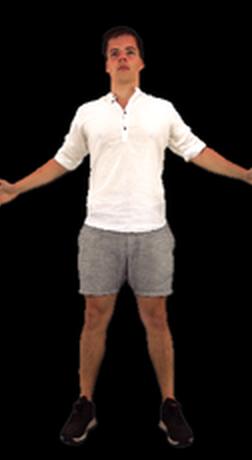}  
      & \includegraphics[width=\imagewidth]{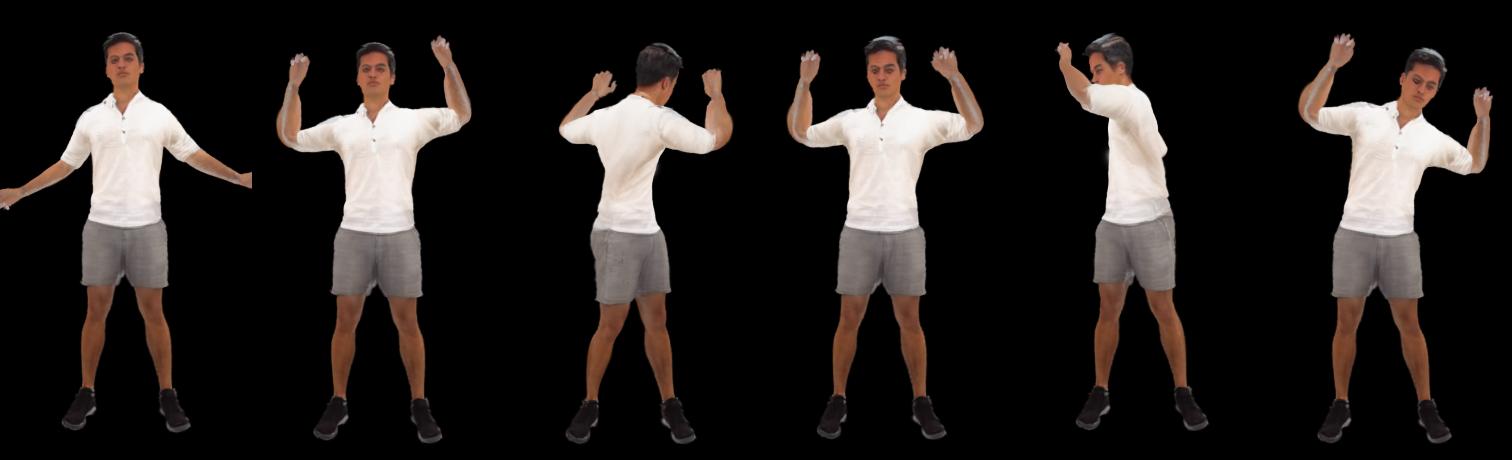} 
      & \includegraphics[width=\imagewidth]{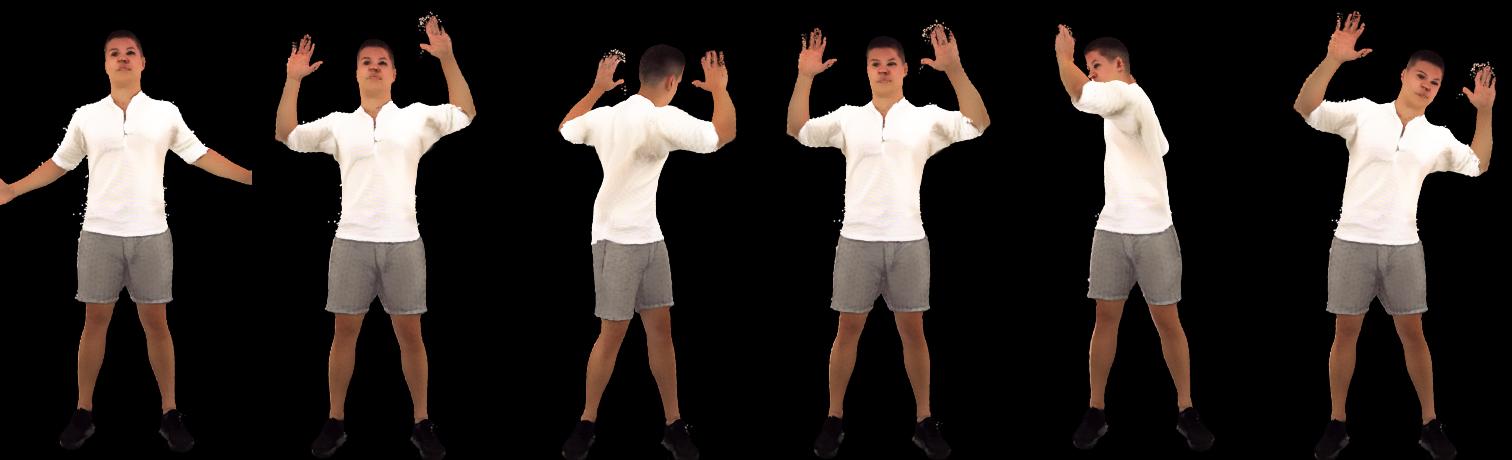} \\
  \includegraphics[width=\inputwidth]{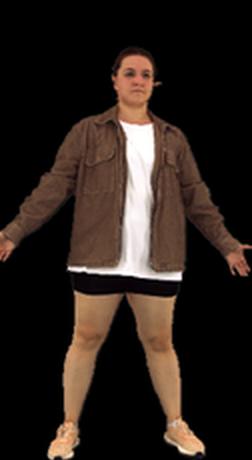} 
      & \includegraphics[width=\imagewidth]{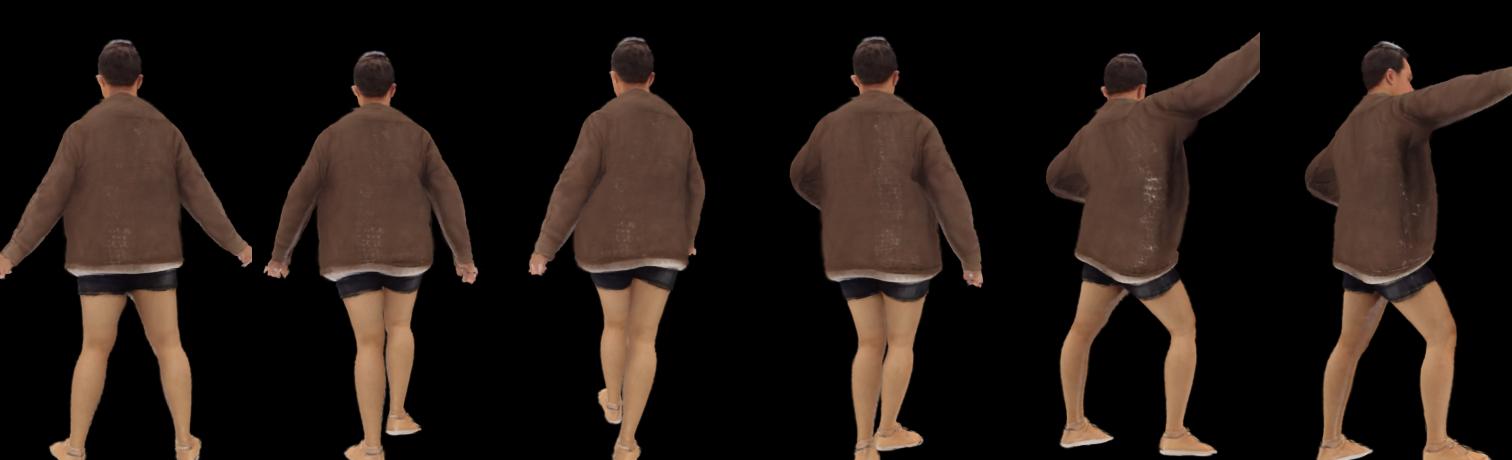}
      & \includegraphics[width=\imagewidth]{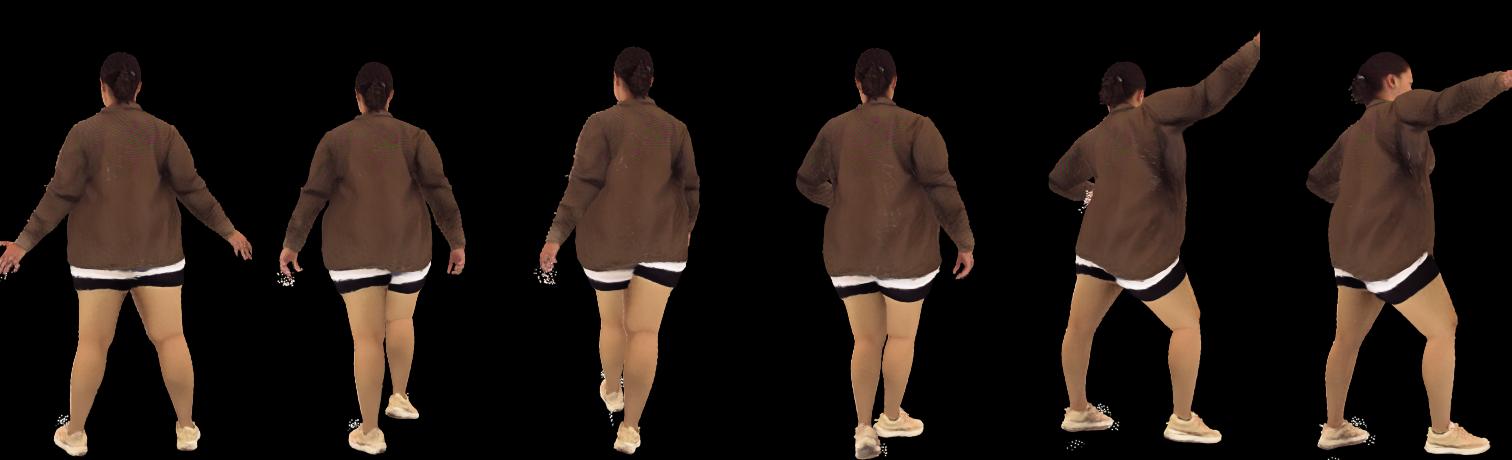} \\
  \includegraphics[width=\inputwidth]{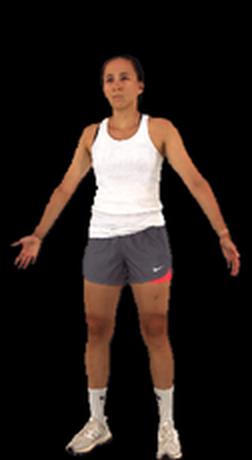}
      & \includegraphics[width=\imagewidth]{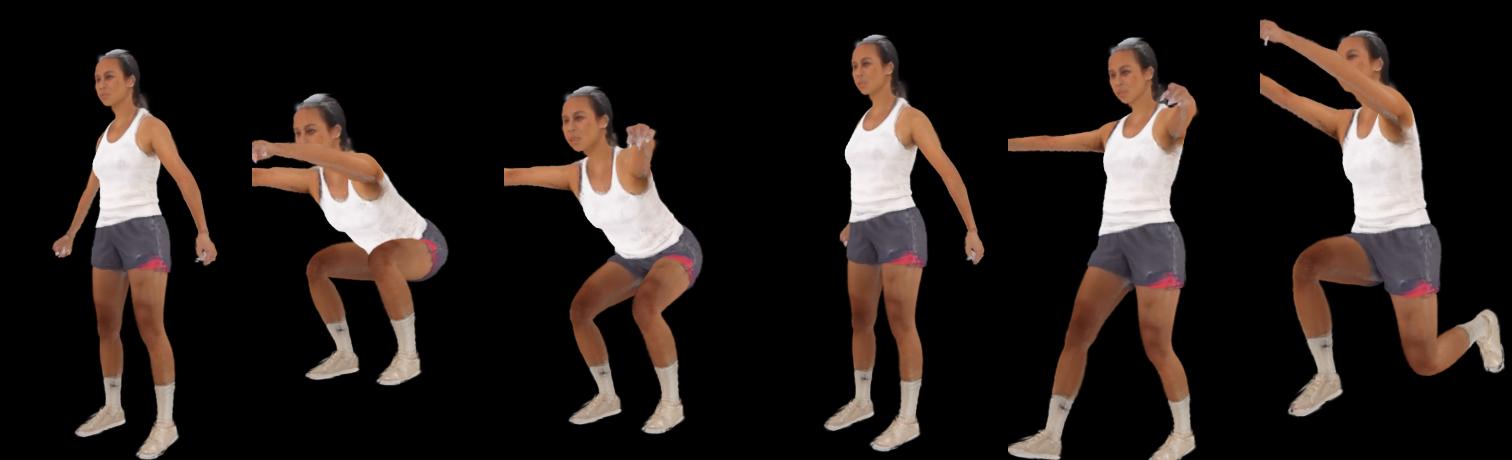} 
      & \includegraphics[width=\imagewidth]{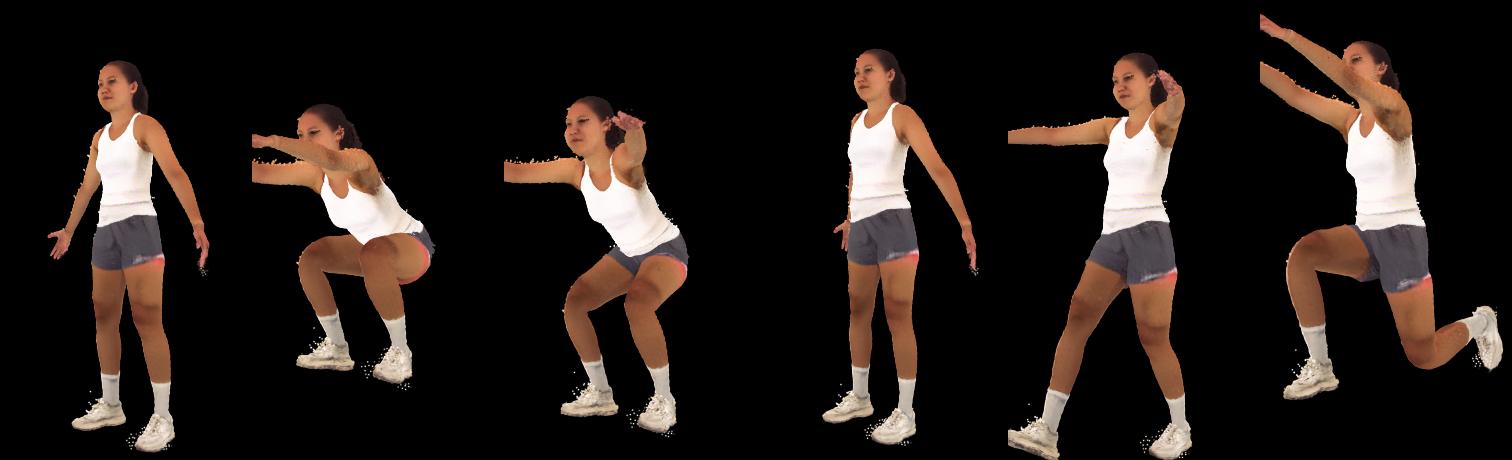} \\
  \includegraphics[width=\inputwidth]{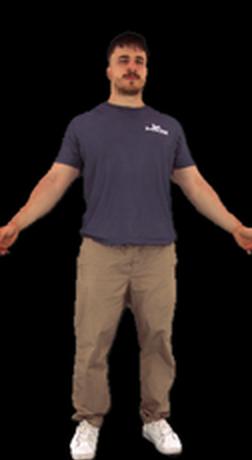}  
      & \includegraphics[width=\imagewidth]{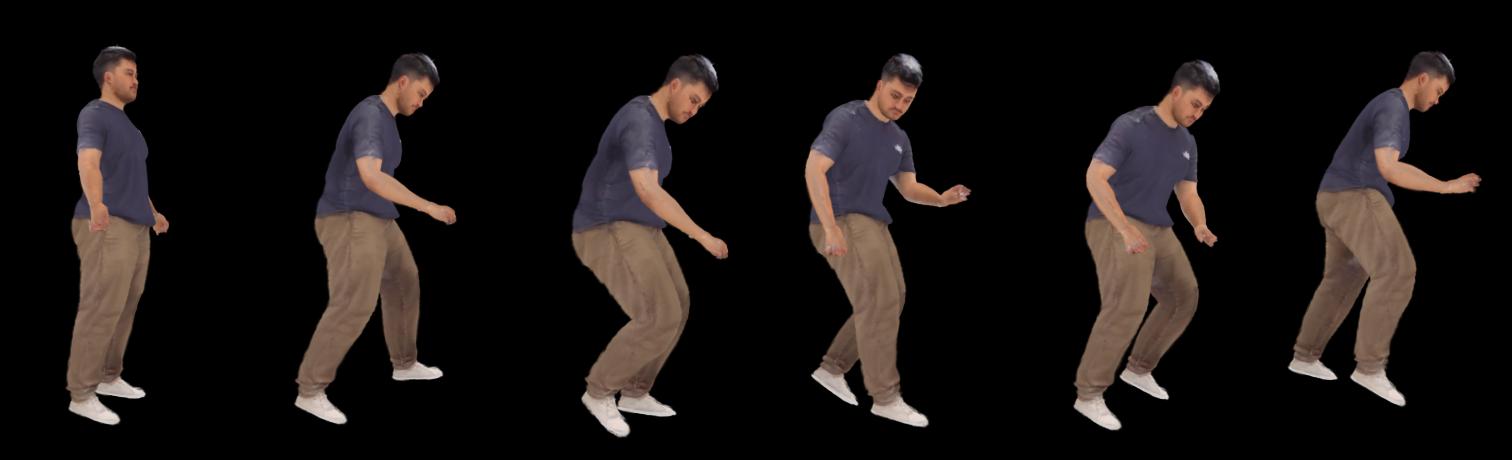} 
      & \includegraphics[width=\imagewidth]{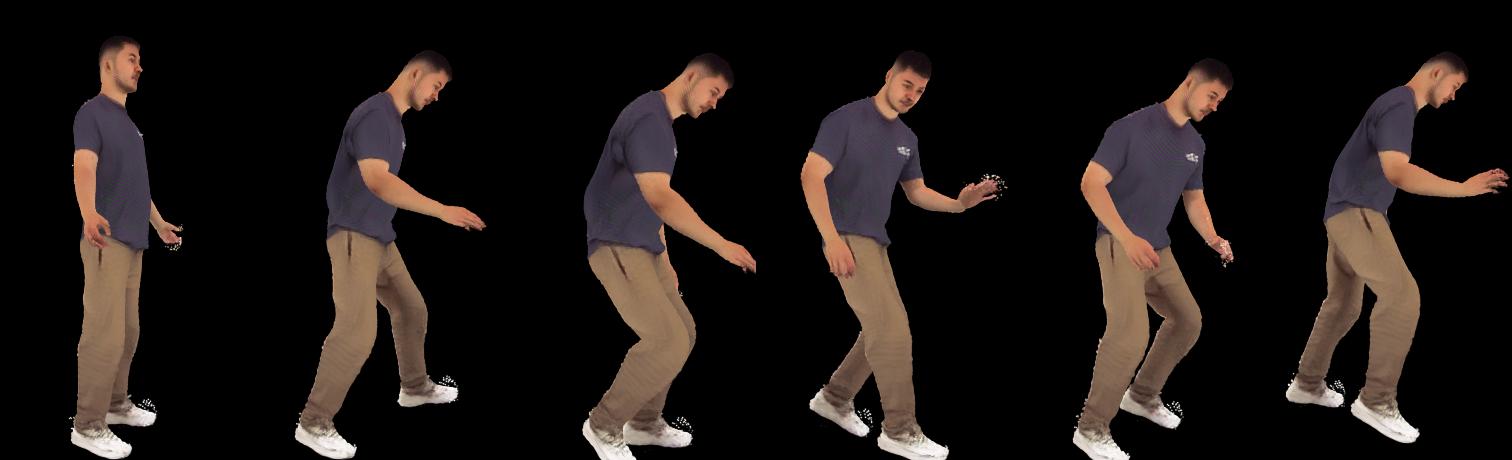} \\
  \includegraphics[width=\inputwidth]{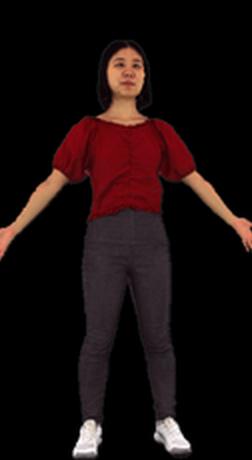}  
      & \includegraphics[width=\imagewidth]{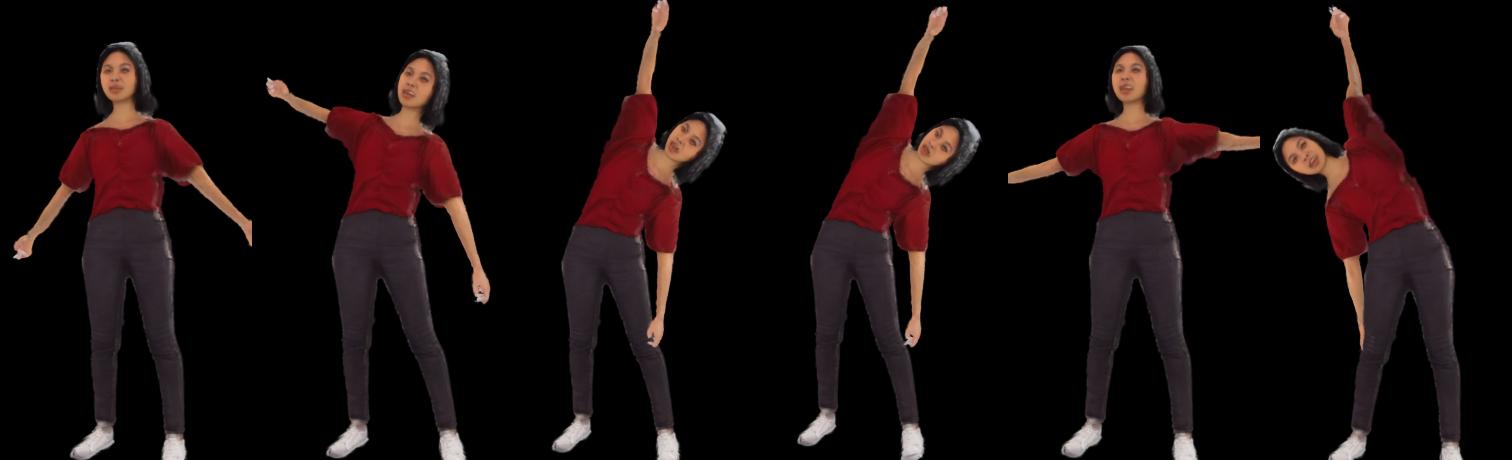} 
      & \includegraphics[width=\imagewidth]{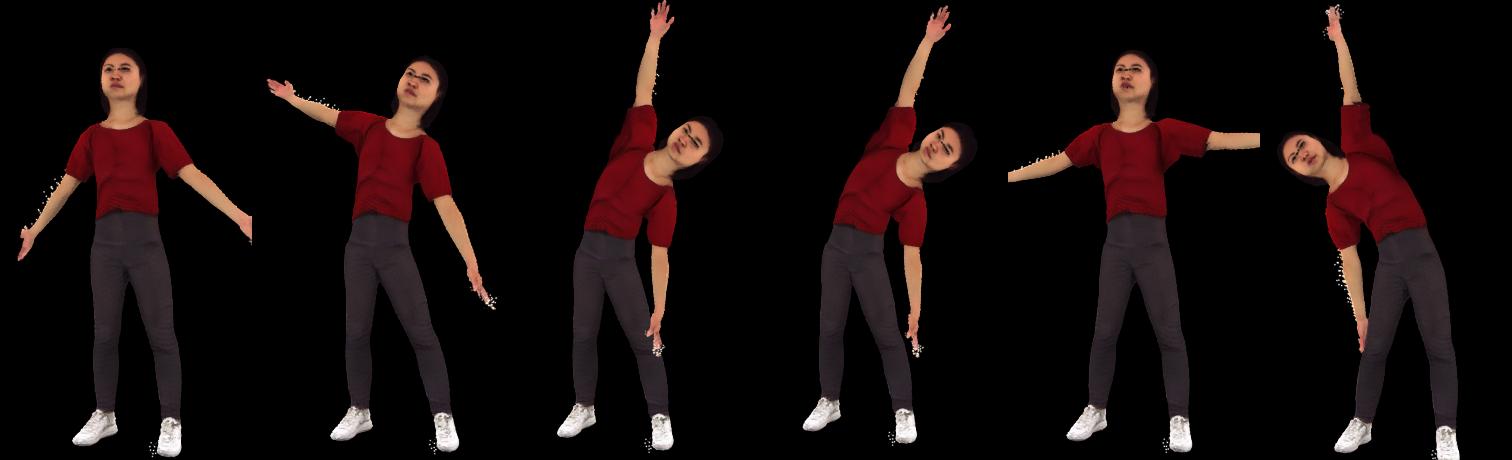} \\
  Input & IDOL~\cite{zhuang2024idolinstantphotorealistic3d} & \textbf{Ours} \\
\end{tabular}
\end{center}
\vspace{-3mm}
\caption{\label{fig:animation_4d}Comparison for animation with IDOL  \cite{zhuang2024idolinstantphotorealistic3d} (concurrent SOTA work). Our Dream, Lift, Animate framework enables detailed renderings for difficult poses, outperforming the state-of-the-art in one-shot animatable avatars in perceptual and photometric metrics. Please see Table \ref{tbl:comp_4d} for metrics.
}
\end{figure*}

\begin{figure*}[ht]
\begin{center}
\includegraphics[width=\textwidth]{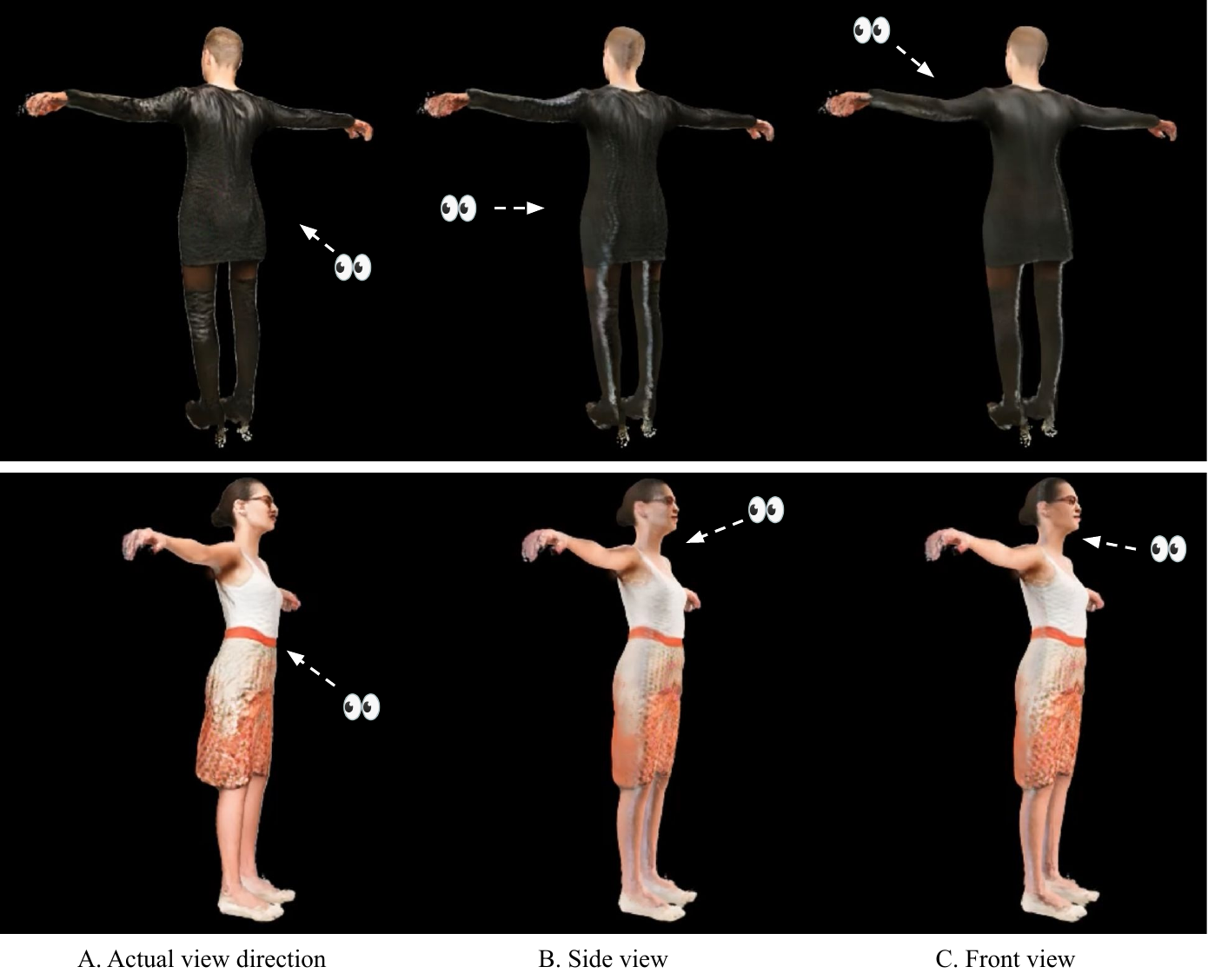} 
\end{center}
\caption{\label{fig:vieweffects}
View-dependent effects.
The Gaussian Parameter Decoder (GPD) conditions on the view direction with Plucker ray maps (Sec. 3.3 and Fig. 4 in the main paper). This enables view-dependent effects.
We render two subjects in an A-pose using the same camera angle but feeding Plucker rays for different view direction: the actual view direction (A), and side and front views (B. and C.). Note the actual view direction (A.) yields sharp details and reflections and other view directions (B. and C.) show effects like a halo on the arms and legs.
}
\end{figure*}

\begin{figure*}[ht]
\begin{center}
\includegraphics[width=0.7\textwidth]{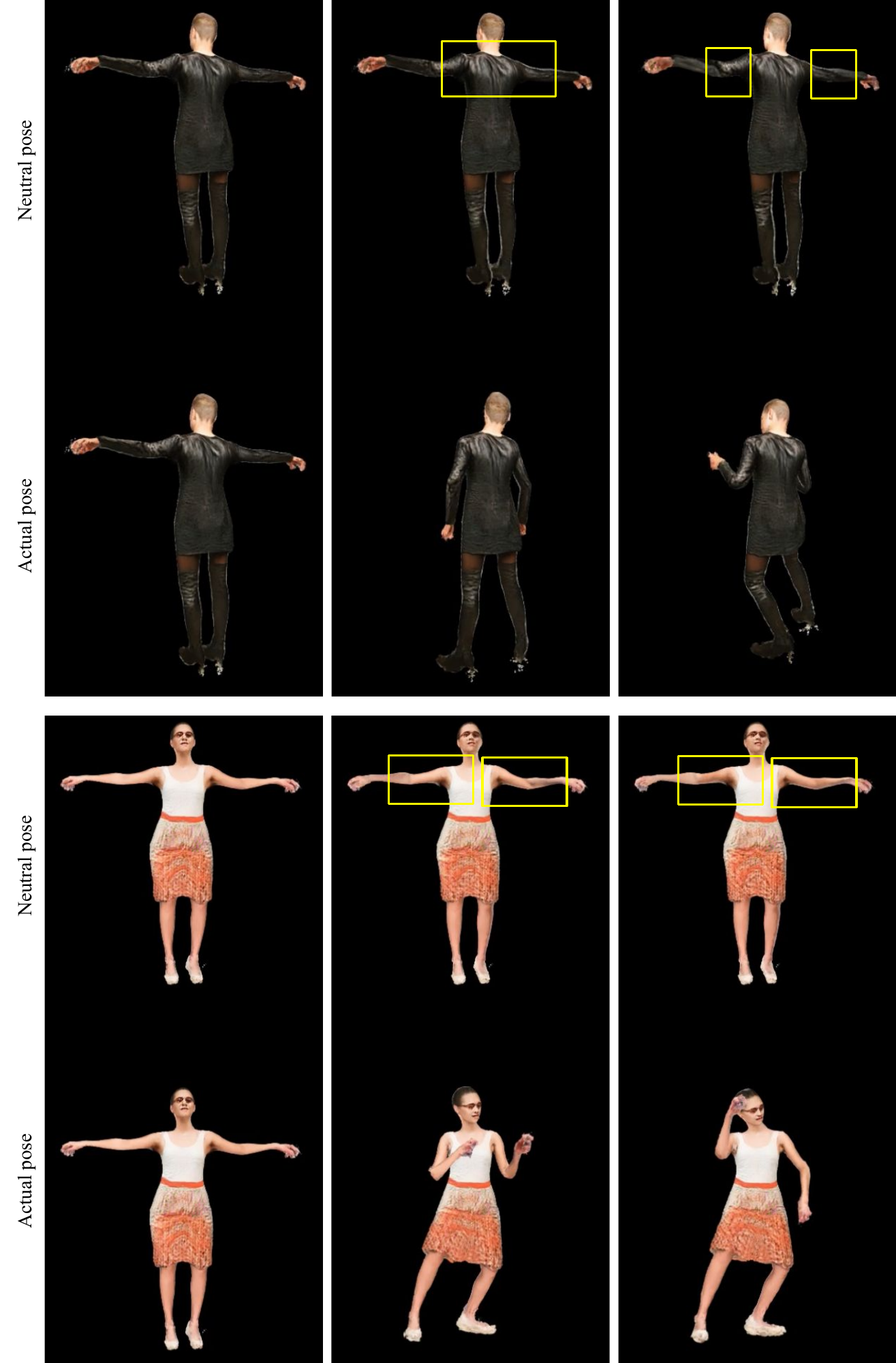} 
\end{center}
\caption{\label{fig:poseeffects}
Pose-dependent effects.
The Gaussian Parameter Decoder (GPD) conditions on the target pose via a relative vertex position map and surface normals (Sec. 3.3 and Fig. 4 in the main paper), enabling pose-dependent effects.
We render two subjects in novel poses and visualize 
the pose-dependent effects, most notable on the shoulders and arms. Please see the website for an animated version.
}
\end{figure*}

\begin{figure*}[ht]
    \centering
  \includegraphics[width=\textwidth]{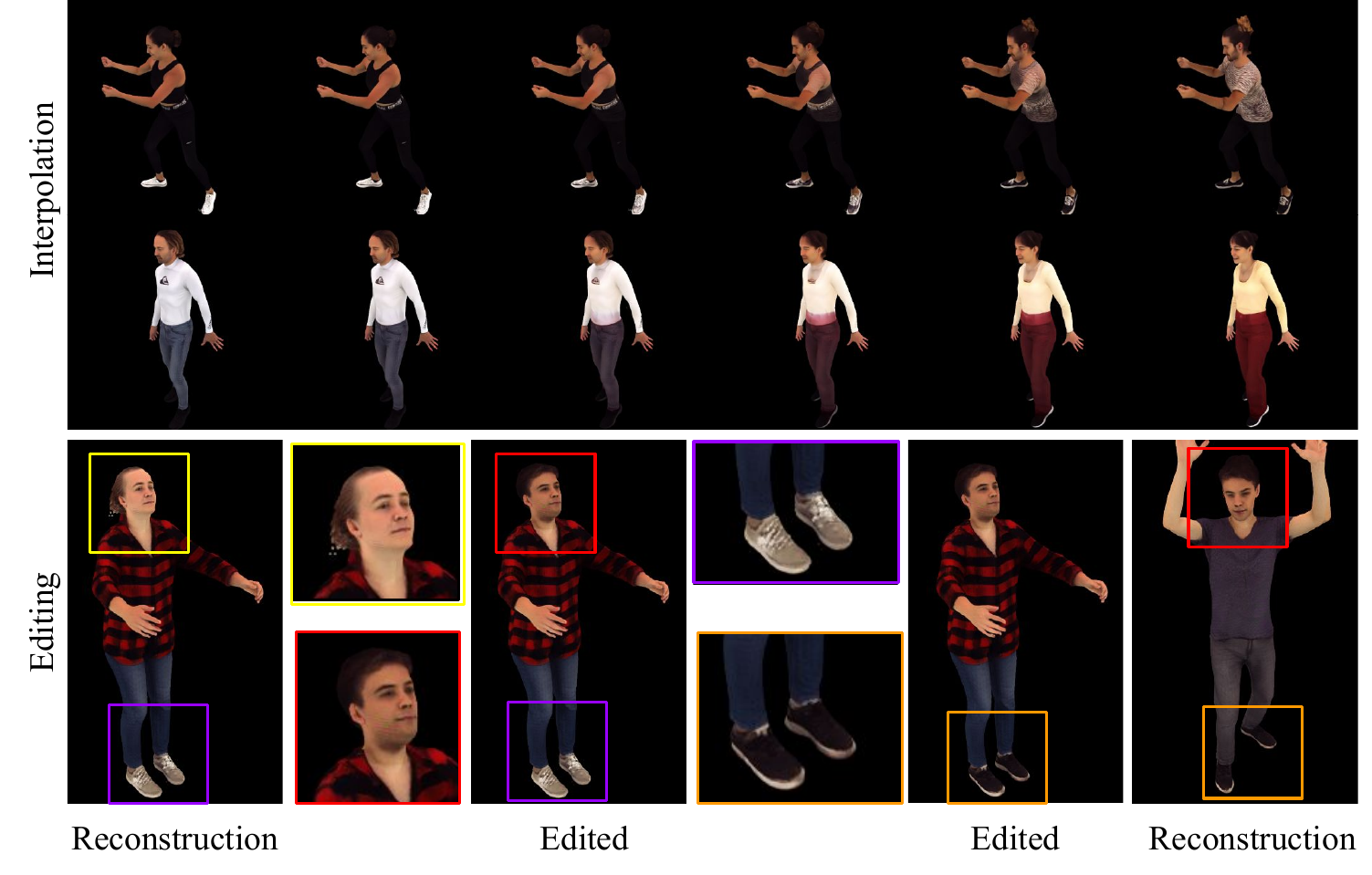}
    \caption{\label{fig:applications}Applications. Our structured latent code affords editing (bottom). In addition, we observe emerging capabilities like smooth interpolations between avatar latent codes. These examples are reconstructions using multi-view inputs from CustomHumans \cite{ho2023custom}.\vspace{-1em}}
\end{figure*}

\subsection{Supplementary Comparisons}
We extend the visual results from the main paper with additional examples from ActorsHQ \cite{isik2023humanrf} in Fig. \ref{fig:comp_ahq} and \ref{fig:animation}, and we provide more examples from in-the-wild SHHQ images in Fig. \ref{fig:shhq}. Tbl. \ref{tbl:comp_ahq} provides supplementary comparisons for novel pose synthesis on ActorsHQ. In addition, we compare with with the best-performing related work, IDOL \cite{zhuang2024idolinstantphotorealistic3d}, on 4D-Dress \cite{wang20244ddress}. Tbl. \ref{tbl:comp_4d} reports metrics, and Fig. \ref{fig:comp_4d} shows visual results for novel view synthesis. We complement these results with a comparison for animation in Fig. \ref{fig:animation_4d}. Finally, Fig. \ref{fig:applications} provides a high-resolution variant of the application figure in the main paper.
 Please see the supplementary video for in-the-wild results on SHHQ \cite{isik2023humanrf}.

\subsection{Supplementary Ablations}
\label{suppsec:ablation}
\begin{figure*}[ht]
\begin{center}
\small
\setlength{\tabcolsep}{2pt}
\newcommand{\inputwidth}{1.4cm}
\begin{tabular}{c cccccccc}
  \includegraphics[width=\inputwidth]{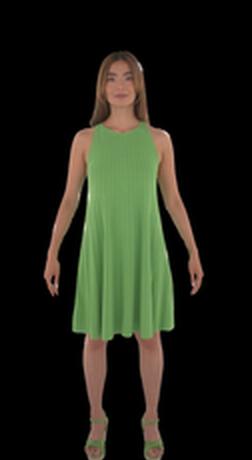}
    &  \includegraphics[width=\inputwidth]{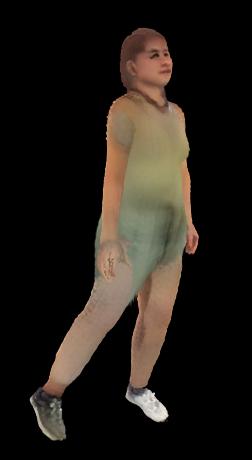}
    &  \includegraphics[width=\inputwidth]{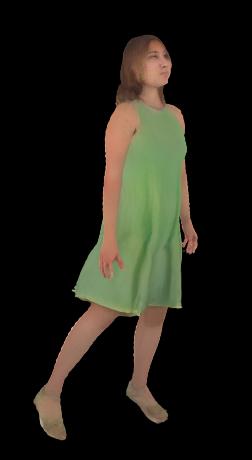}
    &  \includegraphics[width=\inputwidth]{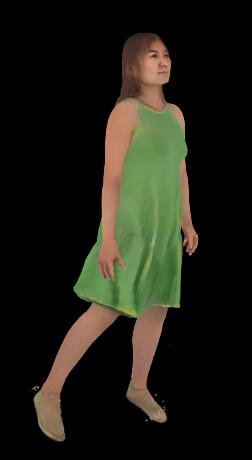}
    &  \includegraphics[width=\inputwidth]{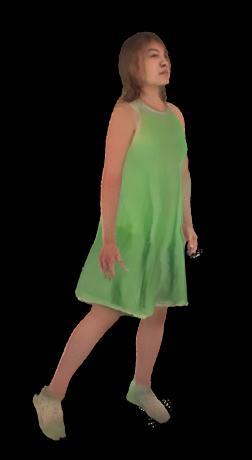}
    &  \includegraphics[width=\inputwidth]{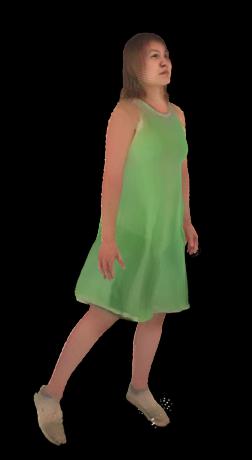}
    &  \includegraphics[width=\inputwidth]{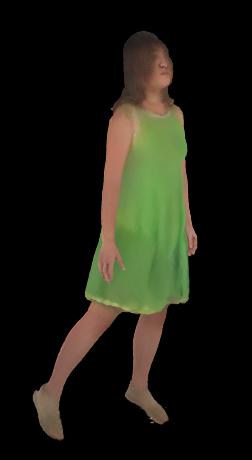}
    &  \includegraphics[width=\inputwidth]{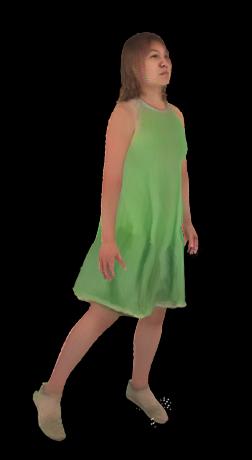}
    &  \includegraphics[width=\inputwidth]{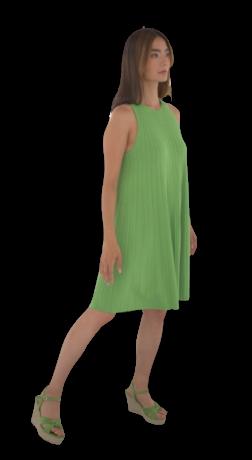} \\
  \includegraphics[width=\inputwidth]{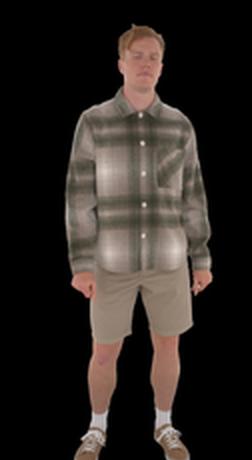}
    &  \includegraphics[width=\inputwidth]{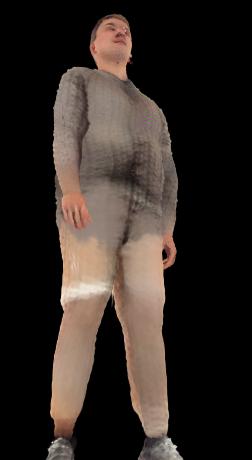}
    &  \includegraphics[width=\inputwidth]{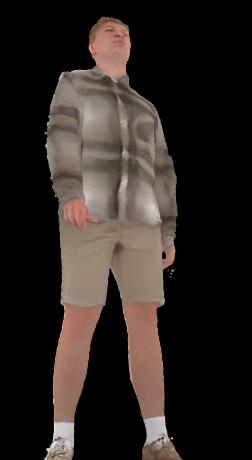}
    &  \includegraphics[width=\inputwidth]{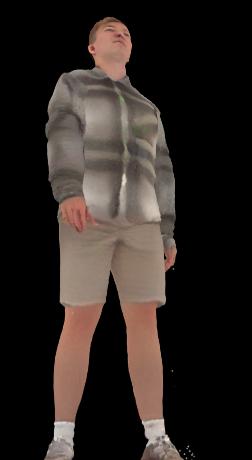}
    &  \includegraphics[width=\inputwidth]{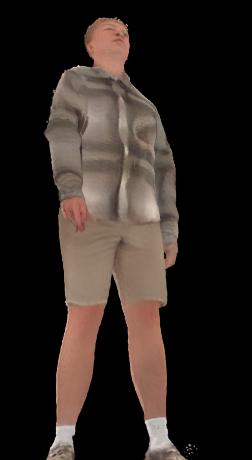}
    &  \includegraphics[width=\inputwidth]{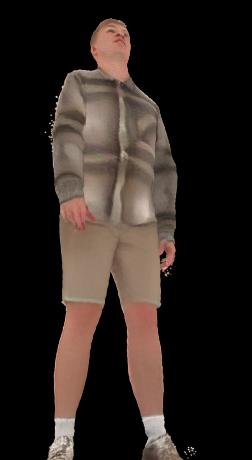}
    &  \includegraphics[width=\inputwidth]{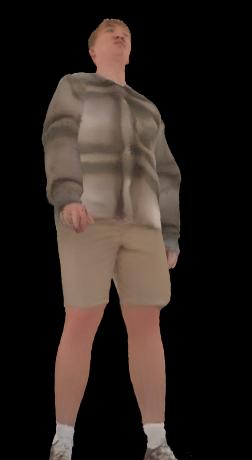}
    &  \includegraphics[width=\inputwidth]{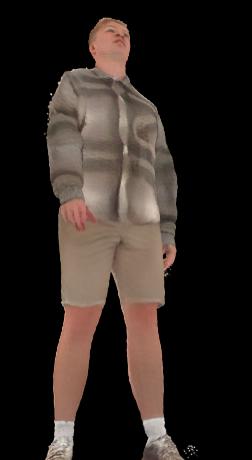}
    &  \includegraphics[width=\inputwidth]{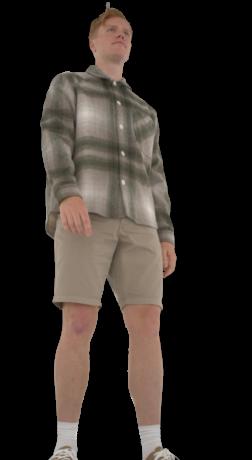} \\
  \includegraphics[width=\inputwidth]{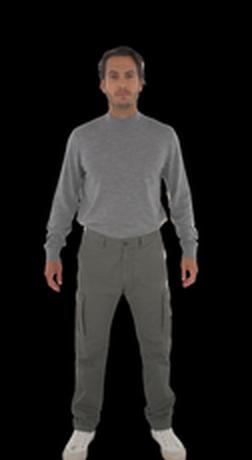}
    &  \includegraphics[width=\inputwidth]{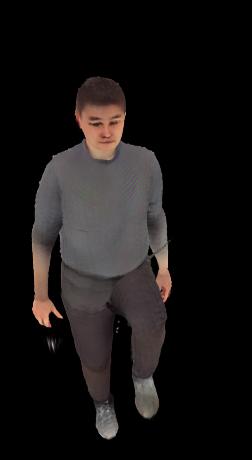}
    &  \includegraphics[width=\inputwidth]{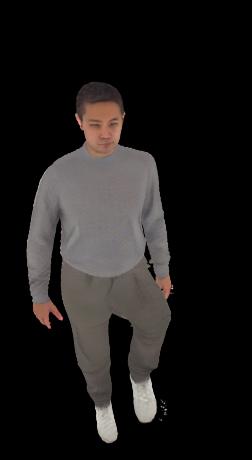}
    &  \includegraphics[width=\inputwidth]{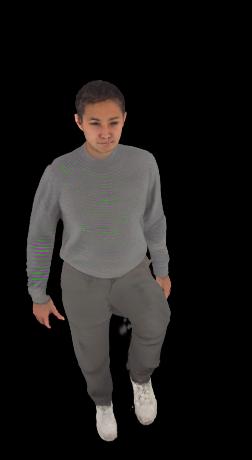}
    &  \includegraphics[width=\inputwidth]{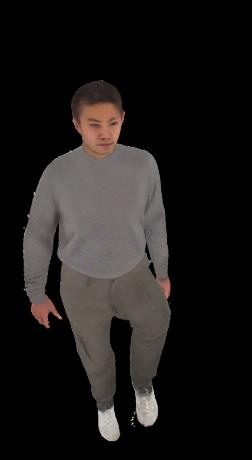}
    &  \includegraphics[width=\inputwidth]{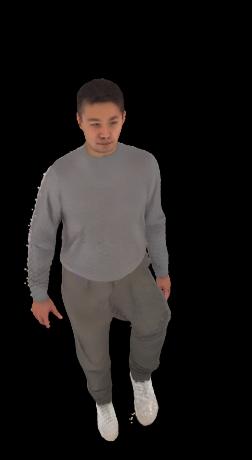}
    &  \includegraphics[width=\inputwidth]{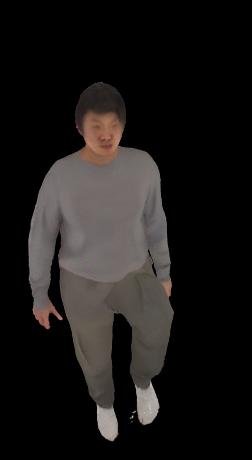}
    &  \includegraphics[width=\inputwidth]{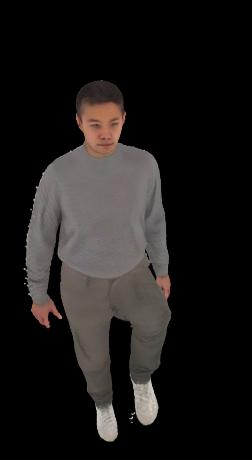}
    &  \includegraphics[width=\inputwidth]{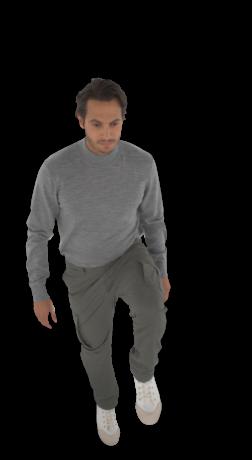} \\
  \includegraphics[width=\inputwidth]{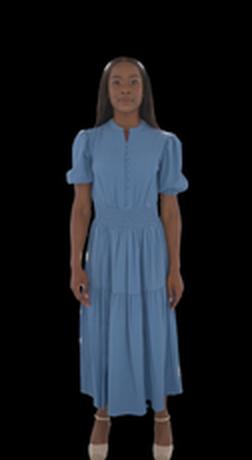}
    &  \includegraphics[width=\inputwidth]{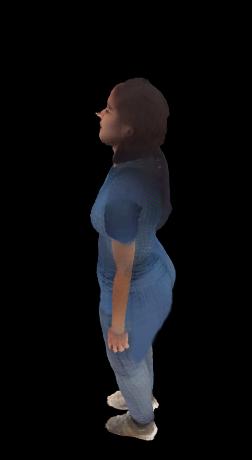}
    &  \includegraphics[width=\inputwidth]{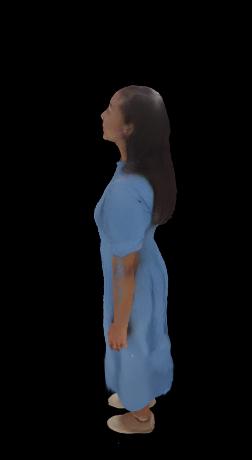}
    &  \includegraphics[width=\inputwidth]{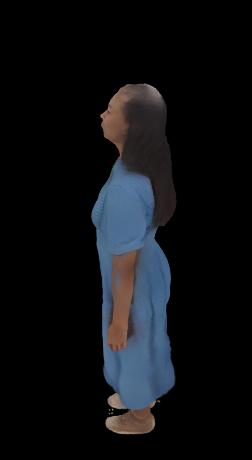}
    &  \includegraphics[width=\inputwidth]{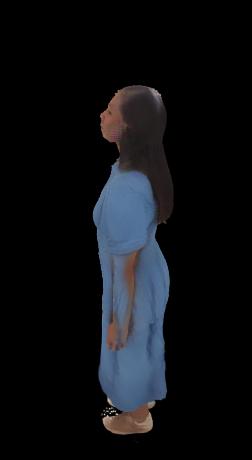}
    &  \includegraphics[width=\inputwidth]{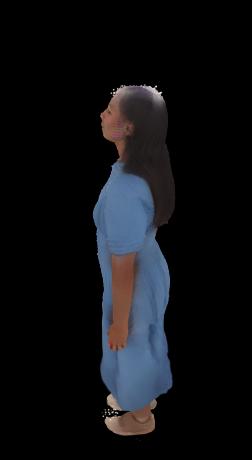}
    &  \includegraphics[width=\inputwidth]{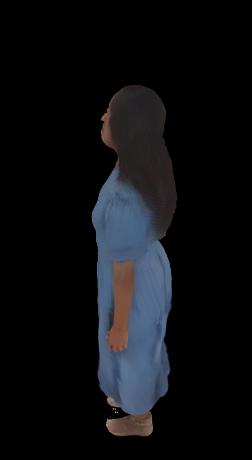}
    &  \includegraphics[width=\inputwidth]{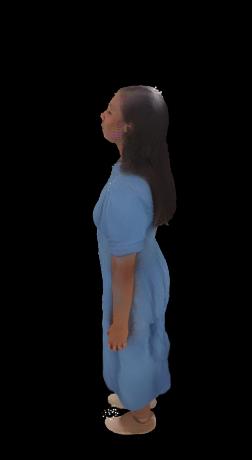}
    &  \includegraphics[width=\inputwidth]{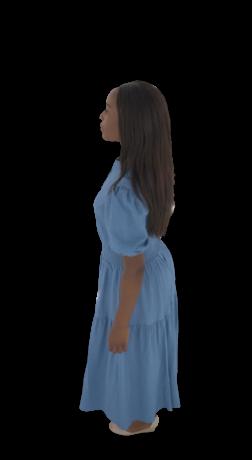} \\
  \includegraphics[width=\inputwidth]{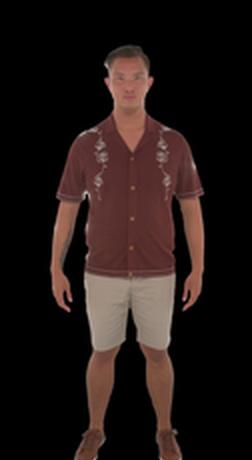}
    &  \includegraphics[width=\inputwidth]{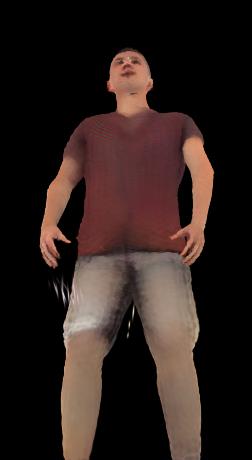}
    &  \includegraphics[width=\inputwidth]{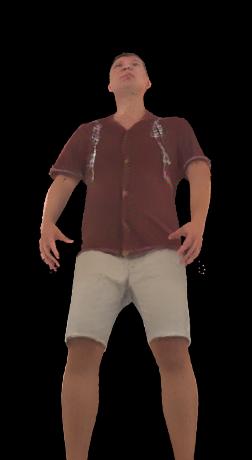}
    &  \includegraphics[width=\inputwidth]{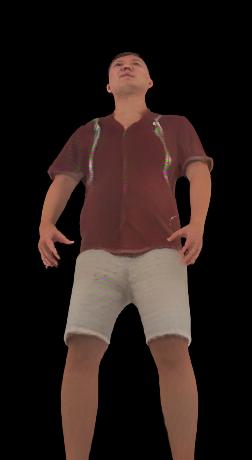}
    &  \includegraphics[width=\inputwidth]{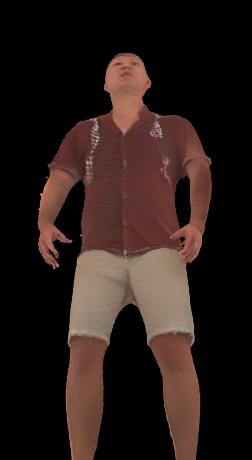}
    &  \includegraphics[width=\inputwidth]{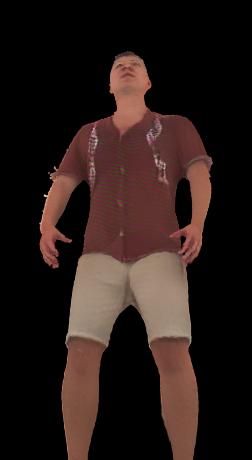}
    &  \includegraphics[width=\inputwidth]{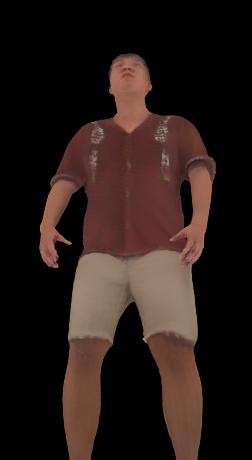}
    &  \includegraphics[width=\inputwidth]{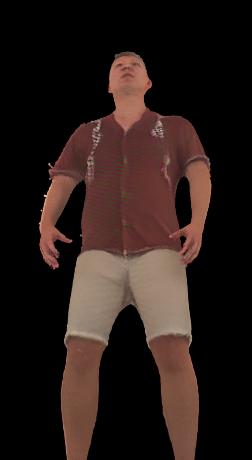}
    &  \includegraphics[width=\inputwidth]{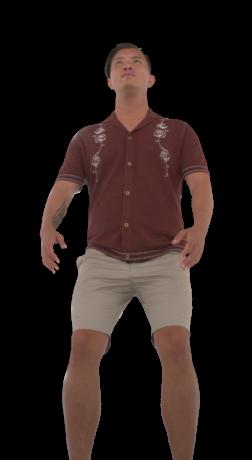} \\
  \includegraphics[width=\inputwidth]{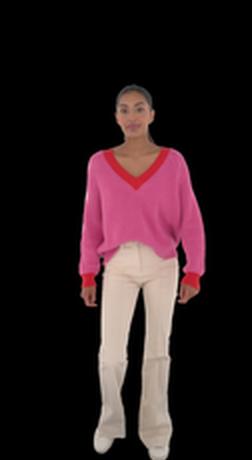}
    &  \includegraphics[width=\inputwidth]{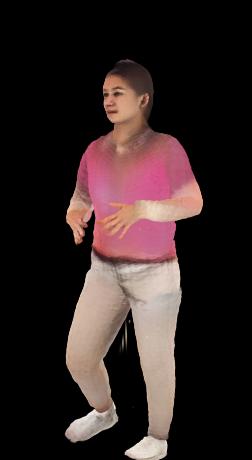}
    &  \includegraphics[width=\inputwidth]{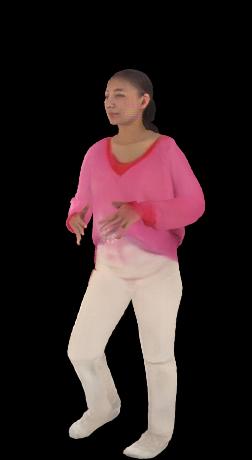}
    &  \includegraphics[width=\inputwidth]{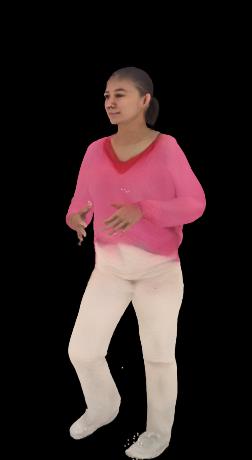}
    &  \includegraphics[width=\inputwidth]{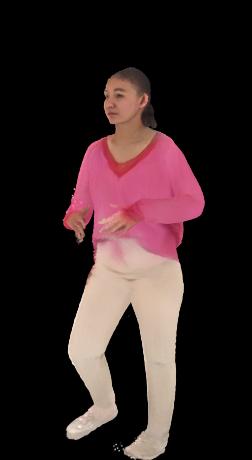}
    &  \includegraphics[width=\inputwidth]{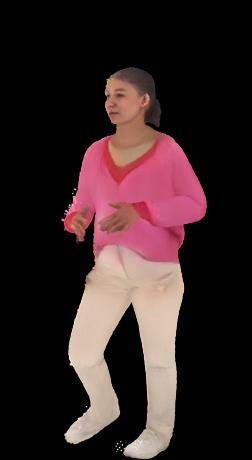}
    &  \includegraphics[width=\inputwidth]{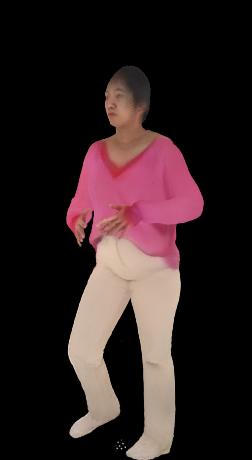}
    &  \includegraphics[width=\inputwidth]{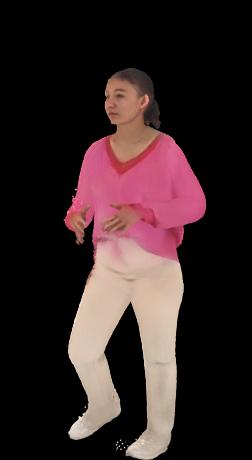}
    &  \includegraphics[width=\inputwidth]{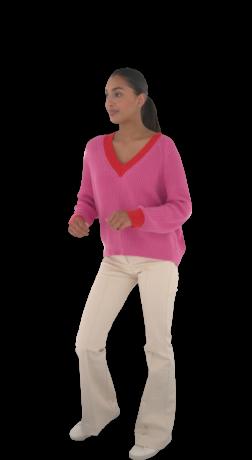} \\
Input &  A.i & A.ii & B. & C.i & C.ii & C.iii  & \textbf{Ours} & GT
\end{tabular}
\end{center}
\caption{\label{fig:ablation}Ablation visuals. A.i shows the results without learning the UV mapping, and A.ii replaces the vertex position map with a learned query, which is optimized during training. B. shows the result when conditionals are missing in the Gaussian Parameter Decoder. C.i feeds noisy SMPL-X parameters, C.ii only synthesized views, and C.iii only ground truth views.
Please see Table \ref{supptbl:ablation} for metrics.
}
\end{figure*}
\begin{table*}[ht]
\footnotesize 
\small
\begin{center}
\begin{tabular}{l ccc ccc ccc}
\hline
    & \multicolumn{3}{c}{\textbf{Novel view}} 
    & \multicolumn{3}{c}{\textbf{Novel pose}}
    & \multicolumn{3}{c}{\textbf{Novel view \& pose}} \\
 \\
    & \textbf{LPIPS} $\downarrow$ & 
    \textbf{PSNR} $\uparrow$ &
    \textbf{SSIM} $\uparrow$
    & \textbf{LPIPS} $\downarrow$ & 
    \textbf{PSNR} $\uparrow$ &
    \textbf{SSIM} $\uparrow$
    & \textbf{LPIPS} $\downarrow$ & 
    \textbf{PSNR} $\uparrow$ &
    \textbf{SSIM} $\uparrow$
    \\
\textbf{A. UV mapping} \\
i) Mesh unprojection (no learning) &0.0779 & 23.78 & 0.9211 & 0.0702 & 24.22 & 0.9273 & 0.0790 & 23.66 & 0.9209\\
ii) Learned query &0.0607 & 25.13 & 0.9283 & 0.0496 & 25.97 & 0.9356 & 0.0652 & 24.67 & 0.9254 \\
\hline
\textbf{B. Animate \& render} \\
No conditioning & 0.0597 & 25.37 & 0.9278 & 0.0502 & 26.16 & 0.9332 & 0.0644 & 24.86 & 0.9248  \\
\hline
\textbf{C. Inputs} \\
i) Noisy SMPL-X inputs & 0.0618 & 25.17 & 0.9262 & 0.0496 & 26.11 & 0.9332 & 0.0660 & 24.77 & 0.9240 \\
ii) 4 synthetic inputs & 0.0625 & 25.13 & 0.9250 & 0.0527 & 25.75 & 0.9307 & 0.0657 & 24.78 & 0.9237
 \\
iii) 4 GT inputs & 0.0566 & 26.40 & 0.9347 & 0.0505 & 26.39 & 0.9358 & 0.0613 & 25.71 & 0.9310
 \\
\hline
\textbf{Ours*} & 0.0592 & 25.54 & 0.9345 & 0.0490 & 26.71 & 0.9418 & 0.0642 & 24.98 & 0.9318 \\
\textbf{Ours} & 0.0580 & 25.58 & 0.9279 & 0.0471 & 26.41 & 0.9351 & 0.0624 & 25.09 & 0.9254 \\
\hline
\end{tabular}
\end{center}
\caption{Supplementary ablation studies (Sec. \ref{suppsec:ablation}). \textbf{Ours*} denotes our model after finetuning on SHHQ.\label{supptbl:ablation}}
\end{table*}

We complement the ablations in the main paper with detailed metrics in Tbl. \ref{supptbl:ablation} and visuals in Fig. \ref{fig:ablation}. The first set of ablations directly projects to the UV space with the SMPL-X mesh without learning the UV mapping (A.i). The second set replaces the vertex position map with a learned query, which is optimized during training (A.ii). When conditionals are missing in the Gaussian Parameter Decoder, the performance degrades, and we observe artifacts close to the surface (B). Finally, we simulate noisy SMLP-X fits (C.i), purely synthesized inputs (C.ii), and the availability of ground truth views (C.iii). The row \textbf{Ours*} indicates our model finetuned on SHHQ.

State-of-the art methods \cite{zhuang2024idolinstantphotorealistic3d,qiu2025LHM} ignore pose-dependent effects and model view-dependent effects with spherical harmonics \cite{kerbl20233d}. 
The design of our Gaussian Parameter Decoder (GPD, Sec. 3.3 and Fig. 4 in the main paper) enables pose- and view-dependent effects by conditioning on a UV-space vertex position map, surface normals, and Plücker rays.
Fig. \ref{fig:poseeffects} visualizes pose-dependent effects by conditioning the Gaussian Parameter Decoder on vertex position maps and surface normals from other samples, and Fig. \ref{fig:vieweffects} shows how the rendering changes when feeding different view directions as conditional inputs to the GPD.

\section{Societal Impact}
\label{suppsec:impact}
While Dream, Lift, Animate (DLA) offers significant societal benefits by making high-quality 3D avatar creation accessible and fostering innovation in digital communication and creativity, it also introduces notable risks. The ability to generate lifelike, animatable avatars from a single image raises concerns about identity misuse, deepfakes, and unauthorized digital replication, which could lead to privacy violations or reputational harm. Additionally, as with all animation technologies, there is a risk of perpetuating stereotypes or misrepresenting cultures if creators do not exercise care and sensitivity in how avatars are depicted. The animation industry has historically faced ethical dilemmas around representation, cultural appropriation, and the portrayal of social groups, making it crucial for researchers to adopt responsible practices that ensure fairness, inclusivity, and respect for diverse identities. Without thoughtful governance and ethical guidelines, the societal impact of such powerful generative tools could skew negative, amplifying biases or enabling exploitation alongside their creative promise.

\end{document}